\documentclass[draft,tightenlines,nofootinbib,preprint,aps,eqsecnum,amsmath,amssymb]{revtex4}

\newcommand{\beq}{\begin{equation}}
\newcommand{\eeq}{\end{equation}}
\newcommand{\bea}{\begin{eqnarray}}
\newcommand{\eea}{\end{eqnarray}}

\begin{document}

\title{
General covariance and the objectivity of space-time
point-events.}

\medskip

\author{Luca Lusanna}

\affiliation{ Sezione INFN di Firenze\\ Polo Scientifico\\ Via Sansone 1\\
50019 Sesto Fiorentino (FI), Italy\\ Phone: 0039-055-4572334\\
FAX: 0039-055-4572364\\ E-mail: lusanna@fi.infn.it}

\author{Massimo Pauri}

\affiliation{
Dipartimento di Fisica - Sezione Teorica\\ Universita' di Parma\\
Parco Area Scienze 7/A\\ 43100 Parma, Italy \\Phone:
0039-0521-905219\\ FAX: 0039-0521-905223\\ E-mail:
pauri@pr.infn.it\\
\ \\
\today\\
 }

\begin{abstract}
\ \\
"The last remnant of physical objectivity of space-time" is
disclosed in the case of a continuous family of spatially
non-compact models of general relativity (GR). The {\it physical
individuation} of point-events is furnished by the intrinsic
degrees of freedom of the gravitational field, (viz, the {\it
Dirac observables}) that represent - as it were - the {\it ontic}
part of the metric field. The physical role of the {\it epistemic}
part (viz. the {\it gauge} variables) is likewise clarified. At
the end the philosophical import of the {\it Hole Argument} is
substantially weakened and in fact the Argument itself dis-solved,
while a peculiar four-dimensional {\it holistic and structuralist}
view of space-time, (called {\it point-structuralism}), emerges,
including elements common to the tradition of both {\it
substantivalism} and {\it relationism}. The observables of our
models undergo real {\it temporal change}: this gives new evidence
to the fact that statements like the {\it frozen-time} character
of evolution, as other ontological claims about GR, are {\it model
dependent}.

\end{abstract}

\maketitle

\vfill\eject

\section{Introduction}

The fact that the requirement of general covariance might involve
a threat to the physical objectivity of the points of space-time
as represented by the theory of gravitation was becoming clear to
Einstein even before the theory he was trying to construct was
completed. It was during the years 1913-1915 that the threat took
form with the famous Hole Argument ({\it Lochbetrachtung})
(Einstein, 1914) \footnote{For a beautiful historical critique see
Norton (1987, 1992, 1993).}. In the literature about classical
field theories space-time points are usually taken to play the
role of individuals, but it is often implicit that they can be
distinguished only by the physical fields they carry. Yet, the
Hole Argument apparently forbids precisely this kind of
individuation, and since the Argument is a direct consequence of
the {\it general covariance} of general relativity (GR), this
conflict eventually led Einstein  to state (our {\it emphasis}):
\begin{quotation}
\noindent That this {\it requirement of general covariance}, which
{\it takes away from space and time the last remnant of physical
objectivity}, is a natural one, will be seen from the following
reflexion... (Einstein, 1916, p.117)
\end{quotation}

Although Einstein quickly bypassed the seeming cogency of the Hole
Argument against the implementation of general covariance on the
purely pragmatic grounds of the so-called {\it Point-Coincidence
Argument}\footnote{The assertion that {\it the reality of the
world-occurrence (in opposition to that dependent on the choice of
reference system) subsists in space-time coincidences.}}, the
issue remained in the background of the theory until the Hole
Argument received new life in recent years with a seminal paper by
John Stachel (1980). This paper, followed seven years later by
Earman and Norton's philosophical argument against the so-called
space-time {\it manifold substantivalism} (Earman \& Norton,
1987), opened a rich philosophical debate that is still alive
today. The Hole Argument was immediately regarded by virtually all
participants in the debate (Bartels, 1984; Butterfield, 1984,
1987, 1988, 1989; Maudlin, 1988; Rynasiewicz, 1994, 1996) as being
intimately tied to the deep nature of space and time, at least as
they are represented by the mathematical models of GR. It must be
acknowledged that until now the debate had a purely philosophical
relevance. From the physicists' point of view, GR has indeed been
immunized against the Hole Argument - leaving aside any underlying
philosophical issue - by simply embodying the Argument in the
statement that {\it mathematically different} solutions of the
Einstein equations related by {\it passive} - as well as {\it
active} (see later) - diffeomorphisms are {\it physically
equivalent}.
\medskip

The main scope of this paper is to show that the immunization
statement quoted above cannot be regarded as the last word on this
matter both from the physical and the philosophical point of view.
From the physical point of view, we will show that physical
equivalence of solutions means much more than mere difference in
mathematical description since it entails equivalence of the
descriptions of phenomena made in different global, non-inertial
frames, which are extended space-time laboratories with their
(dynamically determined) {\it chrono-geometrical conventions} and
inertial potentials. On these same grounds, we will argue from the
philosophical point of view that, first of all, the equivalence
statement cannot have the implications that the Hole Argument has
traditionally attributed to it and, second, that the equivalence
does not compels us to take a definite position in front of the
crude dichotomy {\it substantivalism} versus {\it relationism};
rather we shall show that, for specific models, it leads naturally
to advance a {\it tertium quid} between these two positions that
tries in some sense to overcome the crudeness of the debate by
including elements common to the traditions of both {\it
substantivalism} and {\it relationism}\footnote{The {\it
substantivalist} position is a form of realism about certain
spatiotemporal structures, being committed to believing in the
existence of those entities that are quantified over by our
space-time theories, in particular space-time points. It conceives
space-time, more or less, as a substance, that is as something
that exists independently of any of the things in it. Accordingly,
the extreme form of substantivalism identifies space-time with the
bare mathematical manifold of events, perhaps as mereological
fusion of all the points. On the other hand, the strong {\it
relationist} position is the view that space-time arises as a mere
abstraction from the spatiotemporal properties of other things, so
that spatio-temporal relations are derivative and supervenient on
physical relations obtaining among events and physical objects.
Note that a simple anti-substantivalist position does not deny the
reality of space-time (it is not merely anti-realist) but asserts
that space-time has no reality independently of the bodies of
fields it contains. The crucial question for our notion of
spatiotemporal structuralism, is therefore the specification of
the nature of fields that are {\it indispensable} for the very
definition of physical space-time (e.g. the gravitational field
with its causal structure) as distinguished from {\it other}
physical fields.}. Third, we shall go beyond the simple statement
of equivalence showing that space-time point-events can be {\it
individuated physically} by the intrinsic degrees of freedom of
the gravitational field.

\medskip

It must be clear from the start, however, that, given the enormous
mathematical variety of possible solutions of Einstein's
equations, one should not expect that a clarification of the
possible meaning of a would-be {\it objectivity} of space-time
points could be obtained {\it in general}. More precisely, we
shall indeed conclude that at least three of the main questions we
discuss can be clarified for a definite continuous class of
generic solutions corresponding to spatially non-compact
space-times\footnote{The Christodoulou-Klainermann space-times
(Christodoulou \& Klainermann, 1993).}, but {\it not} for the
spatially compact ones. The former class is also privileged from
the point of view of the inclusion of elementary particles.
Consistently, we do not claim to draw general conclusions about
the ontology of Einstein's space-times. While allowing that the
ontology of general relativistic space-times can be different for
different models, we only claim that a particular class of
solutions exist which naturally leads to a concept of the
space-time ontology consisting in a peculiar form of {\it
structural space-time realism}. Our thesis holds that space-time
point-events (the {\it relata}) do exist and we quantify over
them; their properties are relational being {\it conferred} on
them in a holistic way by the whole structure of the metric field
and the extrinsic curvature on a simultaneity hyper-surface. At
the same time, they are literally {\it identifiable} with the
local values of the intrinsic degrees of freedom of the
gravitational field (Dirac observables), and we will claim that,
in a definite abstract sense, they also possess a special kind of
{\it intrinsic} properties. In this way both the metric field and
the point-events maintain - to paraphrase Newton - their {\it own
manner of existence} and we qualify our conception as
"point-structuralism". Our view does not dissolve physical
entities into mathematical structures, so that it supports a
moderate {\it entity-realist} attitude towards both the metric
field and its point-events, as well as a {\it theory-realist}
attitude towards Einstein's field equations. In conclusion this
work should be considered as a case study for the defence of a
thesis about the nature of the identity of space-time points
characterized by a peculiar form of objectivity. Technically, this
philosophical conclusion is made possible thanks to a specific
methodology of gauge-fixing based on the notion of {\it intrinsic
pseudo-coordinates} introduced by Bergmann and Komar. It should be
stressed that the uniqueness of its mathematical basis (the way in
which the four scalar eigenvalues of the Weyl tensor can be
equated to four {\it scalar radar pseudo-coordinates} by means of
a definite class of gauge-fixing procedures) shows that this
methodology constitutes the only possible way of disclosing the
proper point-events ontology of the class of space-times we are
referring to. For given initial data of the {\it Dirac
observables} (which identify an Einstein's "universe"), any other
kind of gauge-fixing procedure, would lead to gauge-equivalent
solutions in which the underlying point-events ontology simply
would not be shown.

At the technical level, we aim to show that some capabilities
peculiar to the Hamiltonian approach to GR can be exploited for
the purpose of better understanding important interpretive issues
surrounding the theory. The Hamiltonian approach guarantees first
of all that the initial value problem of Einstein's equations is
mathematically well-posed, a circumstance that does not occur in a
natural way within the configurational Lagrangian framework
(Friedrich \& Rendall, 2000; Rendall, 1998). Furthermore, on the
basis of the Shanmugadhasan canonical transformation
(Shanmugadhasan, 1973; Lusanna, 1993), this framework provides a
net distinction between {\it physical observables} (the four
so-called {\it Dirac observables}) connected to the (two)
intrinsic degrees of freedom of the gravitational field, on one
hand, and {\it gauge variables}, on the other. The latter - which
express the typical arbitrariness of the theory and must be fixed
({\it gauge-fixing}) before solving the Einstein equations for the
intrinsic degrees of freedom - turn out to play a fundamental
role, no less than the {\it Dirac observables}, in clarifying the
real import of the Hole Argument. It will be seen indeed that the
resulting {\it gauge} character of GR is a crucial factor not only
for clarifying the issue of the objectivity of space-time points,
but also for a true dis-solution of the Hole Argument as to its
philosophical implications\footnote{Our stance about the content
and the implications of the original Hole Argument contrasts with
the manifestly covariant and generalized attitude towards the Hole
phenomenology expounded by John Stachel in many papers (see e.g.
Stachel \& Iftime, 2005, and references therein). We shall defend
our approach in Sections III and VI.}. We stress that reaching
these conclusions within the Lagrangian formulation would be
technically quite awkward if not impossible, since the Legendre
pull-back of the non-point canonical transformations of the
Hamiltonian formulation requires tools like the infinite-jet
bundle formalism.

\medskip

For the above reasons, the discussion in the following sections
will be substantially grounded upon the fact that GR is a gauge
theory, i.e., a theory in which the physical system being dealt
with is described by more variables than there are physically
independent degrees of freedom. Such extra variables are
introduced to make the description more transparent and
mathematically handy but require correspondingly a {\it gauge
symmetry} having the role of extracting the physically relevant
content. The important fact in our case is that while, from the
mathematical point of view of the constrained Hamiltonian
formalism, GR is a theory like any other (e.g., electromagnetism
and Yang-Mills theory), from the {\it physical} point of view it
is {\it radically different}, just because of its gauge invariance
under a group of diffeomorphisms acting on space-time itself,
instead of invariance under the action of a local {\it inner} Lie
group. Furthermore, in GR we cannot rely from the beginning on
empirically validated, gauge-invariant dynamical equations for the
{\it local} fields, as it happens with electro-magnetism, where
Maxwell equations can be written in terms of the gauge invariant
electric and magnetic fields. On the contrary, Einstein's general
covariance (viz. the gauge freedom of GR) is such that the
introduction of extra ({\it gauge}) variables does indeed make the
{\it mathematical} description of general relativity more {\it
transparent} (through manifest general covariance instead of
manifest Lorentz covariance) but, at least {\it prima facie}, by
ruling out any background structure at the outset, it also makes
its physical interpretation more intriguing, and conceals at the
same time the intrinsic properties of point-events. In GR the
distinction between what is observable and what is not, is
unavoidably entangled with the constitution of the very {\it
stage}, space--time, where the play of physics is enacted: a
stage, however, which also takes an active part in the play. In
other words, the gauge-fixing mechanism plays the {\it dual role}
of making the dynamics {\it unique} (as in all gauge theories),
and of fixing the {\it appearance} of the spatio-temporal
dynamical background. At the same time, this mechanism highlights
a characteristic functional split of the metric tensor that can be
briefly described as follows. First of all, any gauge-fixing is
equivalent to the constitution of a {\it global, non-inertial,
extended, space-time laboratory with its coordinates, for every
$\tau$}, as well as to a {\it dynamical determination} of the {\it
conventions} about {\it distant simultaneity}. In particular,
different {\it conventions} within the same space-time (the same
{\it "universe"}), turn out to be simply {\it gauge-related
options}. Therefore, on one hand, the Dirac observables specify -
as it were - the {\it ontic} structure of space-time connected to
the intrinsic degrees of freedom of the gravitational field (and -
physically - to {\it tidal-like} effects). On the other, the gauge
variables specify the built-in {\it epistemic}
component\footnote{In this paper {\it ontic} refers to the
essential content of the gravitational field {\it qua} physical
entity with two independent degrees of freedom, {\it epistemic} to
that part of the information encoded in the metric that must be
freely specified in order to get empirical access to the essential
part (it refers therefore to the {it\ appearance} of the
gravitational phenomena). We are perfectly aware that we are here
overstating the philosophical import of terms like {\it ontic} and
{\it epistemic} and their relationships. Nothing, however, hinges
on these nuances in what follows.} of the metric tensor
(physically related to the {\it generalized inertial effects}
accessible in each extended laboratory)\footnote{Such a global,
extended laboratory is a non-rigid, non-inertial frame (the only
existing in GR) centered on the (in general) accelerated observer
whose world-line is the origin of the 3-coordinates (Lusanna \&
Pauri, 2004a): hereafter it will be denoted by the acronym NIF.
Any NIF is the result of a {\it complete gauge-fixing} (see
Section IV). The gauge-fixing procedure determines the {\it
appearance of phenomena} by determining uniquely the form of the
inertial forces (Coriolis, Jacobi, centrifugal,...) in each point
of a NIF. A crucial difference of this mechanism in GR with
respect to the Newtonian case is the fact that the inertial
potentials depend upon {\it tidal effects}, besides the
coordinates of the non-inertial frame.}

Summarizing, the gauge variables play a multiple role in
completing the structural properties of the general-relativistic
space-time: their fixing is necessary for solving Einstein's
equations, for reconstructing the four-dimensional chrono-geometry
emerging from the initial values of the {\it four Dirac
observables} and for allowing empirical access to the theory
through the definition of a spatiotemporal {\it laboratory}. It's
important to stress in this connection that in GR, unlike in
ordinary gauge theories, the {\it reduced phase space ${\tilde
\Omega}_4$} of the abstract observables (a realization of which is
created by every gauge fixing) plays an abstract role, since the
inertial effects associated to each NIF are lost because of the
quotient procedure (see Sections IV,V,VI).

\bigskip

Apart from the dis-solution of the Hole Argument, which is
expounded in Section III, the main result of our analysis is given
in Section V where we show how the {\it ontic} part of the metric
(the intrinsic degrees of freedom of the gravitational field) may
confer a {\it physical individuation} onto space-time
points\footnote{There is an unfortunate ambiguity in the usage of
the term {\it space-time points} in the literature: sometimes it
refers to elements of the mathematical structure that is the first
layer of the space-time model, and sometimes to the points
interpreted as {\it physical} events. We will adopt the term {\it
point--event} in the latter sense and simply {\it point} in the
former.}. Since - as will be seen - such degrees of freedom depend
in a highly {\it non-local} way upon the values of the metric and
the extrinsic curvature over a whole space-like surface of distant
simultaneity, point-events receive a peculiar sort of {\it
properties} that are conferred on them {\it holistically} by the
whole simultaneous metrical structure. Admittedly, the distinction
between {\it ontic} and {\it epistemic} parts, as well as the {\it
form} of the space-like surfaces of distant simultaneity, are
NIF-dependent\footnote{Yet, according to a {\it main conjecture}
we have advanced elsewhere (see Lusanna \& Pauri, 2004a, 2004b), a
canonical basis of {\it scalars} (coordinate-independent
quantities), or at least a Poisson algebra of them, should exist,
making the above distinction between Dirac observables and gauge
variables {\it fully invariant}, see Section VI and footnote 32.}.

Finally, an additional important feature of the solutions of GR
dealt with in our discussion is the following. The ADM formalism
(Arnowitt \& Deser \& Misner, 1962) for {\it spatially compact}
space-times without boundary implies that the Dirac Hamiltonian
generates purely harmless gauge transformations, so that, being
zero on the reduced phase space, it {\it cannot engender any real
temporal change}. This is the origin of the so-called {\it frozen
evolution} description; in this connection see Earman (2002),
Belot \& Earman  (1999, 2001). However, in the case of the
Christodoulou - Klainermann continuous family of {\it spatially
non-compact} space-times, {\it internal mathematical consistency}
(requiring the addition of the DeWitt surface term to the
Hamiltonian (DeWitt, 1967), see later) entails that the generator
of temporal evolution, namely the (now non-weakly vanishing) Dirac
Hamiltonian, be instead the so-called {\it weak ADM energy}. This
quantity {\it does generate real temporal modifications} of the
canonical variables. In conclusion, this shows again that also
statements like the assertion of the {\it frozen-time} picture of
evolution, as other ontological claims about GR, are {\it model
dependent}.

The main part of the technical developments underlying this work
have already been introduced elsewhere (Pauri \& Vallisneri, 2002,
Lusanna \& Pauri, 2004a, 2004b;) where additional properties of
the Christodoulou-Klainermann family of space-times are also
discussed. For a more general philosophical presentation, see
Dorato \& Pauri (2004).

\section{Noether and dynamical symmetries}

Standard general covariance, which essentially amounts to the
statement that the Einstein equations for the metric field
${}^4g(x)$ have a tensor character, implies first of all that the
basic equations are form invariant under general coordinate
transformations ({\it passive} diffeomorphisms), so that the
Lagrangian density in the Einstein-Hilbert action is singular.
Namely, passive diffeomorphisms {\it are local Noether symmetries}
of the action, so that Dirac constraints appear correspondingly in
the Hamiltonian formulation. The singular nature of the
variational principle of the action entails in turn that four of
the Einstein equations be in fact {\it Lagrangian constraints},
namely restrictions on the Cauchy data, while four combinations of
Einstein's equations and their gradients vanish identically ({\it
contracted Bianchi identities}). Thus, the ten components of the
solution ${}^4g_{\mu\nu}(x)$ are in fact functionals of only two
"deterministic" dynamical degrees of freedom and eight further
degrees of freedom which are left {\it completely undetermined} by
Einstein's equations {\it even once the Lagrangian constraints are
satisfied}. This state of affairs makes the treatment of both the
Cauchy problem of the non-hyperbolic system of Einstein's
equations and the definition of observables within the Lagrangian
context (Friedrich \& Rendall, 2000; Rendall, 1998) extremely
complicated.

For the above reasons, standard general covariance is then
interpreted, in modern terminology, as the statement that {\it a
physical solution of Einstein's equations} properly corresponds to
a {\it 4-geometry}, namely the equivalence class of all the
4-metric tensors, solutions of the equations, written in all
possible 4-coordinate systems. This equivalence class is usually
represented by the quotient ${}^4Geom = {}^4Riem / {}_PDiff\,
M^4$, where ${}^4Riem$ denotes the space of metric tensor
solutions of Einstein's equations and ${}_PDiff\ $ is the infinite
group of {\it passive} diffeomorphisms (general coordinate
transformations). On the other hand, any two {\it inequivalent}
Einstein space-times are different 4-geometries or "universes".

Consider now the abstract differential-geometric concept of {\it
active} diffeomorphism $D_A$ and its consequent action on the
tensor fields defined on the differentiable manifold $M^4$ [see,
for example, (Wald, 1984, pp.438-439)]. An {\it active}
diffeomorphism $D_A$ maps points of $M^4$ to points of $M^4$:
$D_A: p {\rightarrow \hspace{.2cm}}\ p' = D_A \cdot p$. Its
tangent map $D_A^{*}$ maps tensor fields $T {\rightarrow
\hspace{.2cm}} D_A{*} \cdot T$ in such a way that $[T](p)
{\rightarrow \hspace{.2cm}} [D_A^{*} \cdot T](p) \equiv
[T^{'}](p)$. Then $[D_A^{*} \cdot T](p) = [T](D_A^{-1}\cdot p)$.
It is seen that the transformed tensor field $D_A^{*} \cdot T$ is
a {\it new} tensor field whose components in general will have at
$p$ values that are {\it different} from those of the components
of $T$. On the other hand, the components of $D_A^* \cdot T$ have
at $p'$ - by construction - the same values that the components of
the original tensor field $T$ have at $p$: $T^{'}(D_A \cdot p) =
T(p)$ or $T'(p) = T(D_A^{-1}\cdot p)$. The new tensor field $D_A^*
\cdot T$ is called the {\it drag-along} (or {\it push-forward}) of
$T$. There is another, non-geometrical - so-called {\it dual} -
way of looking at the active diffeomorphisms (Norton, 1987). This
{\it duality} is based on the circumstance that in each region of
$M^4$ covered by two or more charts there is a one-to-one
correspondence between an {\it active} diffeomorphism and a
specific coordinate transformation. The coordinate transformation
${\cal T}_{D_A}: x(p) {\rightarrow \hspace{.2cm}}\ x'(p) = [{\cal
T}_{D_A}x](p)$ which is {\it dual} to the active diffeomorphism
$D_A$ is defined so that $[{\cal T}_{D_A}x](D_A \cdot p) = x(p)$.
Essentially, this {\it duality} transfers the functional
dependence of the new tensor field in the new coordinate system to
the old system of coordinates. By analogy, the coordinates of the
new system $[x']$ are said to have been {\it dragged-along} with
the {\it active} diffeomorphism $D_A$. It is important to note
here, however, that the above {\it dual view} of active
diffeomorphisms, as particular {\it coordinate}-transformations,
is defined for the moment only {\it implicitly}.

In the abstract coordinate-independent language of differential
geometry, Einstein's equations for the vacuum

\bea
 {}^4G_{\mu\nu}(x)\,\, {\buildrel {def}\over =}\,\, {}^4R_{\mu\nu}(x)
- {1\over 2}\, {}^4R(x)\, {}^4g_{\mu\nu}(x)  = 0.
 \label{E1}
  \eea

\noindent can be written as $G = 0$, where $G$  is the Einstein
2-tensor ($G = G_{\mu\nu}(x)\, dx^{\mu} \bigotimes dx^{\nu}$ in
the coordinate chart $x^{\mu}$). Under an {\it active
diffeomorphism} $D_A: M^4 \mapsto M^4$, $D_A \in {}_ADiff\, M^4$,
we have $G = 0 \mapsto D^*_A\, G = 0$, which shows that active
diffeomorphisms are {\it dynamical symmetries} of the Einstein's
tensor equations, i.e., {\it they map solutions into solutions}.

We have clarified elsewhere (Lusanna \& Pauri, 2004a) the explicit
relationships\footnote{At least for the infinitesimal active
transformations.} existing between passive and active
diffeomorphisms on the basis of an important paper by Bergmann and
Komar (1972) in which it is shown that {\it the largest group of
passive dynamical symmetries of Einstein's equations} is not
${}_PDiff\, M^4$ [$x^{{'}\, \mu} = f^{\mu}(x^{\nu})$] but instead
a larger group of transformations of the form

\bea
 Q:&& x^{{'}\, \mu} = f^{\mu}(x^{\nu}, {}^4g_{\alpha\beta}(x)),
 \nonumber \\
 &&{}\nonumber \\
  {}^4g^{'}_{\mu\nu}(x^{'}(x)) &=& {{\partial h^{\alpha}(x^{'},
  {}^4g^{'}(x^{'}))}\over {\partial
 x^{'\, \mu}}}\, {{\partial h^{\beta}(x^{'}, {}^4g^{'}(x^{'}))}\over
 {\partial x^{'\, \nu}}}\, {}^4g_{\alpha\beta}(x).
 \label{E2}
 \eea
\medskip

In the case of completely Liouville-integrable systems, dynamical
symmetries can be re-interpreted as maps of the space of Cauchy
data onto itself. Although we don't have a general proof of the
integrability of Einstein's equations, we know that if the initial
value problem is well-posed and defined\footnote{It is important
to stress that in looking for global solutions of Einstein's
equations as a system of partial differential equations, a number
of preliminary specifications must be given. Among other things:
a) the topology of space-time; b) whether the space-time is
spatially-compact or asymptotically flat at spatial infinity; c)
whether or not in the spatially-compact case there is a spatial
boundary; d) the nature of the function space and the class of
boundary conditions, either at spatial infinity or on the spatial
boundary, for the 4-metric and its derivatives (only in the
spatially-compact case without boundary there is no need of
boundary conditions, replaced by periodicity conditions, so that
these models of GR show the well-known Machian aspects which
influenced Einstein and Wheeler). After these specifications have
been made, a {\it model} of GR is identified. What remains to be
worked out is the characterization of a well-posed initial value
problem. Modulo technicalities, this requires choosing a
4-coordinate system and finding which combinations of the
equations are of {\it elliptic} type (restrictions on the Cauchy
data) and which are of {\it hyperbolic} type (evolution
equations), namely the only ones requiring an initial value
problem. At the Hamiltonian level, the elliptic equations are the
first-class constraints identifying the constraint sub-manifold of
phase space (see Section IV), while the hyperbolic equations are
the Hamilton equations in a fixed gauge (a completely fixed
Hamiltonian gauge corresponds {\it on-shell} to a 4-coordinate
system, see Section IV). When the gauge variables can be separated
from the Dirac observables, only the latter need an initial value
problem (the gauge variables are {\it arbitrary}, modulo
restrictions upon their range coming from the structure of the
gauge orbits inside the constraint sub-manifold). Finally, given a
space-like Cauchy surface in a 4-coordinate system (or in a fixed
Hamiltonian gauge), {\it each admissible set of Cauchy data gives
rise to a different "universe" with the given boundary
conditions}. Clearly, each universe is defined modulo passive
diffeomorphisms changing both the 4-coordinate system and the
Cauchy surface (or modulo the Hamiltonian gauge transformations
changing the gauge and the Cauchy surface) and also modulo the
({\it on-shell}) active diffeomorphisms.}, as it is in the ADM
Hamiltonian description, {\it the space of Cauchy data is
partitioned in gauge-equivalent classes of data}: all of the
Cauchy data in a given class identify a single 4-geometry or {\it
"universe"}. Therefore, under the given hypothesis, {\it the
dynamical symmetries of Einstein's equations fall in two classes
only}: a) those mapping different {\it "universes"} among
themselves, and b) those acting within a single Einstein {\it
"universe"}, mapping gauge-equivalent Cauchy data among
themselves. It is remarkable that, at least for the subset $Q'
\subset Q$ (passive counterpart of a subset ${}_ADiff^{'}\, M^4
\subset {}_ADiff\, M^4$) that corresponds to {\it mappings among
gauge-equivalent Cauchy data}, the transformed metrics do indeed
belong to the {\it same 4-geometry}, i.e. the same equivalence
class generated by applying all {\it passive diffeomorphisms} to
the original 4-metrics: $ {}^4Geom  = {}^4Riem / {}_PDiff\, M^4 =
{}^4Riem / Q'$\footnote{Note, incidentally, that this circumstance
is mathematically possible only because ${}_PDiff\, M^4$ is a {\it
non-normal dense} sub-group of $Q^{'}$.}.

\medskip

Note finally that: a) an {\it explicit} passive representation of
the infinite group of ${}_ADiff\, M^4$ is necessary anyway for our
Hamiltonian treatment of the Hole Argument as well as for any
comparison of the various viewpoints existing in the literature
concerning the {\it solutions} of Einstein's equations; b) the
group $Q'$ describes the dynamical symmetries of Einstein's
equations which are {\it also} local Noether symmetries of the
Einstein-Hilbert action. The 4-metrics reached by using passive
diffeomorphisms are, as it were, only a dense sub-set of the
metrics obtainable by means of the group Q.

\medskip

In conclusion, what is known as a {\it 4-geometry}, is also an
equivalence class of solutions of Einstein's equations {\it
modulo} the subset of dynamical symmetry transformations
${}_ADiff'\, M^4$, whose passive counterpart is $Q^{'}$.
Therefore, following Bergmann \& Komar (1972), Wald (1984), we can
state \footnote{Eqs.(2.3) are usually taken for granted in
mathematical physics, at least at the heuristic level. Since,
however, the control in large of the group manifold of
infinite-dimensional groups like ${}_PDiff\, M^4$ and $Q^{'}$ is,
as yet, an open mathematical issue, one cannot be more rigorous on
this point: see also the end of Section III. For more details
about these issues, the interested reader should see Lusanna \&
Pauri (2004a, 2004b) where a new subset ${}Q_{can}$ of $Q$ is
introduced, namely the Legendre pullback of the {\it on-shell}
Hamiltonian canonical transformations. We distinguish {\it
off-shell} considerations, made within the variational framework
before restricting to the dynamical solutions, from {\it on-shell}
considerations, made after such a restriction. In (Lusanna \&
Pauri, 2004a and 2004b) it is shown, for instance, that we have
also $ {}^4Geom = {}^4Riem / Q_{can}$, since, modulo
technicalities, we have ${}Q_{can} = Q'$. Note that ${}_PDiff\,
M^4 \cap Q_{can}$ are the {\it passive diffeomorphisms which are
re-interpretable as Hamiltonian gauge transformations}.}

\bea
 {}^4Geom = {}^4Riem / {}_PDiff\, M^4 = {}^4Riem / Q' = {}^4Riem /
 {}_ADiff'\, M^4.
 \label{E3}
 \eea

\section{The Hole Argument and its dis-solution}

Although the issue could not be completely clear to Einstein in
1916, as shown by Norton  (1987, 1992, 1993), it is precisely the
nature of dynamical symmetry of the {\it active diffeomorphisms}
that has been considered as expressing the {\it physically
relevant} content of {\it general covariance}, as we shall
presently see.

Remember, first of all that a {\it mathematical model} of GR is
specified by a four-di\-men\-sio\-nal mathematical manifold $M^4$
and by a metrical tensor field $g$, where the latter  represents
{\it both} the chrono-geometrical structure of space-time {\it
and} the potential for the inertial-gravitational field.
Non-gravitational physical fields, when they are present, are also
described by dynamical tensor fields, which appear to be sources
of the Einstein equations. Assume now that $M^4$ contains a {\it
hole} $\mathcal{H}$: that is, an open region where all the
non-gravitational fields vanish so that {\it the metric obeys the
homogeneous Einstein equations}. On $M^4$ we can define an {\it
active} diffeomorphism $D^{*}_{A}$ that re-maps the points inside
$\mathcal{H}$, but blends smoothly into the identity map outside
$\mathcal{H}$ and on the boundary. By construction, for any point
$x \in \mathcal{H}$ we have (in the abstract tensor notation)
$g'(D_A x)= g(x)$, but of course $g'(x) \neq g(x)$ (in the same
notation). The crucial fact is that from the general covariance of
Einstein's equations it follows that if $g$ is one of their
solutions, so is the {\it drag-along} field $g' \equiv D^{*}_{A}
g$.

What is the correct interpretation of the new field $g'$? Clearly,
the transformation involves an {\it active redistribution of the
metric over the points of the manifold in $\mathcal{H}$}, so the
critical question is whether and how the points of the manifold
are primarily {\it individuated}. Now, {\it if} we think of the
points of $\mathcal{H}$ as {\it intrinsically} {\it individuated}
{\it physical events}, where {\it intrinsic} means that their
identity is autonomous and independent of any physical field, the
metric in particular - a claim that is associated with any kind of
{\it manifold substantivalism} - then $g$ and $g'$ must be
regarded as {\it physically distinct} solutions of the Einstein
equations (after all, $g'(x) \neq g(x)$ at the {\it same} point
$x$). This appears as a devastating conclusion for the causality,
or better, the {\it determinateness}\footnote{We prefer to avoid
the term {\it determinism}, because we believe that its
metaphysical flavor tends to overstate the issue at stake. This is
especially true if {\it determinism} is taken in opposition to
{\it indeterminism}, which is not mere absence of {\it
determinism}. We are concerned here with the question of being
{\it determinate} or {\it under-determinate}, referred to
solutions of Einstein's equations.} of the theory, because it
implies that, even after we specify a physical solution for the
gravitational and non-gravitational fields outside the hole - in
particular, {\it on a Cauchy surface for the initial value
problem} - we are still unable to predict a unique physical
solution within the hole.

According to Earman and Norton (1987), the way out of the Hole
Argument lies in abandoning {\it manifold substantivalism}: they
claim that if diffeomorphical\-ly-related metric fields were to
represent different physically possible "universes", then GR would
turn into an {\it indeterministic} theory. And since the issue of
whether determinism holds or not at the {\it physical} level
cannot be decided by opting for a {\it metaphysical} doctrine like
{\it manifold substantivalism}, they conclude that one should
refute any kind of such {\it substantivalism}. Since, however,
relationism does not amount to the mere negation of
substantivalism, and since the literature contains so many
conflicting usages of the term "relationism", they do not simply
conclude that space-time is relational. They state the more
general assumption (which - they claim - is applicable to all
space-time theories) that "diffeomorphic models in a space-time
theory represent the same physical situation". i.e. must be
interpreted as describing the same {\it "universe"} ({\it Leibniz
equivalence}).

The fact that the Leibniz equivalence seems here no more than a
sophisticated re-phrasing of what physicists consider a foregone
conclusion for general relativity, should not be taken at face
value, for the real question for the opposing "sensible
substantivalist" is whether or not space-time should be simply
identified with the bare manifold deprived of any physical field,
and of {\it the metric field} in particular, as Earman and Norton
do, instead of with a set of points each endowed with its own {\it
metrical fingerprint}\footnote{See, for example, Bartels (1994)
and Maudlin (1988).}. Actually, this substantivalist could sustain
the conviction - as we ourselves do - that the metric field,
because of its basic {\it causal structure}, has ontological
priority (Pauri, 1996) over all other fields and, therefore, it is
{\it not} like any other field, as Earman and Norton would have
it. And, according to this view, as we shall  presently see,
Leibniz equivalence is not the last word on the issue
\footnote{Let us recall that already in 1984 Michael Friedman was
lucidly aware of the unsatisfactory status of the understanding of
the relation between diffeomorphic models in terms of Leibniz
equivalence, when he wrote ({\it our emphasis}) "Further, if the
above models are indeed equivalent representations of the same
situation (as it would seem they must do) then {\it how do we
describe this physical situation {\it intrinsically}?}. Finding
such an {\it intrinsic} characterization (avoiding quantification
over {\it bare} points) appears to be a non-trivial, {\it and so
far unsolved mathematical problem}. (Note that it will not do
simply to replace points with equivalence classes of points: for,
in many cases, the equivalence class in question will contain {\it
all} points of the manifold)", see Friedman, (1984).}.

We do believe that the bare manifold of points, deprived of the
infinitesimal pythagorean structure defining the basic distinction
between temporal and spatial directions, let alone the causal
structure which teach all the other fields how to move, can hardly
be seen as space-time. Consequently, and in agreement with Stachel
(1993), we believe that asserting that $g$ and $D^{*}_{A} g$
represent {\it one and the same gravitational field} implies that
{\it the mathematical individuation} of the points of the
differentiable manifold by their coordinates {\it has no physical
content until a metric tensor is specified} \footnote{{\it
Coordinatization} is the only way to individuate the points {\it
mathematically} since, as stressed by Hermann Weyl: ''There is no
distinguishing objective property by which one could tell apart
one point from all others in a homogeneous space: at this level,
fixation of a point is possible only by a {\it demonstrative act}
as indicated by terms like {\it this} and {\it there}.'' (Weyl.
1946, p. 13). See also Schlick, (1917), quoted in M.Friedman,
(2003), p.165.}. Stachel stresses that if $g$ and $D^{*}_{A}g$
must represent the same gravitational field, they cannot be
physically distinguished in any way. Accordingly, when we act on
$g$ with $D^{*}_{A}$ to create the {\it drag-along} field
$D^{*}_{A} g $, no element of physical significance can be left
behind: in particular, {\it nothing} that could identify a point
$x$ of the manifold itself as the {\it same point} of space-time
for both $g$ and $D^{*}_{A} g$. Instead, when $x$ is mapped onto
$x' = D^{*}_{A}x$, it {\it carries over its identity}, as
specified by $g'(x')= g(x)$. This means,  for one thing, that "the
last remnant of physical objectivity" of space-time points, if
any, should be sought for in the physical content of the metric
field itself.

These remarks led Stachel to the important conclusion that {\it
vis \'{a} vis} the physical point-events, the metric actually
plays the role of {\it individuating field}. Precisely, Stachel
suggested that this individuating role could be implemented by
four invariant functionals of the metric, which Komar and Bergmann
(Komar 1958; Bergmann \& Komar, 1960) had already considered.
Stachel, however, did not follow up on this proposal by providing
a concrete realization in terms of solutions of Einstein's
equations, something that we instead will presently do. At the
same time we will show in Section VI that Stachel's suggestion, as
it stands, remains at a too abstract level and fails to exploit
the crucial distinction between {\it ontic} and arbitrary {\it
epistemic} content of the Bergmann-Komar invariant functionals of
the metric, that is necessary to specify the solutions they apply
to.

\medskip

We conclude this Section by summarizing the implications of our
analysis about the meaning and the philosophical import of the
Hole Argument. The force of the {\it indeterminacy argument}
apparently rests on the following basic facts: (i) a solution of
Einstein's equations must be {\it preliminarily} individuated
outside (and, of course, inside) the Hole, otherwise there would
be no meat for the Argument itself. Although the original
formulation of the Hole Argument, as well as many subsequent
expositions of it, are silent on this point, we will see that the
Hole Argument is unavoidably entangled with the initial value
problem \footnote{It is interesting to find that David Hilbert
stressed this point already in 1917 (Hilbert, 1917).}; (ii) the
{\it active} diffeomorphism $D^{*}_{A}$, which is purportedly
chosen to be the {\it identity} outside the {\it hole}
$\mathcal{H}$, is a {\it dynamical symmetry} of Einstein's
equations, so that it maps solutions into solutions, equivalent
(as 4-geometries or Einstein "universes") or not ; (iii) since
$D^{*}_{A}$ is, by hypothesis, the identity on the Cauchy
hyper-surface, it cannot map a solution defining a given Einstein
"universe" into a different "universe", {\it which would
necessarily correspond to inequivalent Cauchy data}; but (iv)
nevertheless, we are told by the Hole Argument that $D^{*}_{A}$
engenders a "different" solution inside the Hole.

Actually, in spite of the {\it prima facie} geometric obviousness
of the identity condition required for $D^{*}_{A}$ outside the
Hole, it is quite illusory trying to explain all the facets of the
relations of the Argument with the initial value problem in the
purely abstract way of differential geometry. The point is that
the differential-geometric 4-D formulation cannot exploit the
advantage that the Hamiltonian formulation possess of working {\it
off-shell} (i.e. before going onto solutions of the Hamilton
equations: see footnote 13). This is crucial since in the present
context the 4-D active diffeomorphisms - {\it qua} dynamical
symmetries of Einstein's equations - must be directly applied on
{\it solutions} of GR. These solutions, however, cannot be
exaustively managed in the 4-D configurational approach in terms
of initial data because of the non-hyperbolicity of Einstein's
equations. The Hole Argument needs that the Cauchy problem be
formulated outside the Hole explicitly and {\it in advance}, a
fact that requires abandoning the Lagrangian way in favor of
Hamiltonian methods. At this point, the results of the previous
Section (the passive counterpart of $D^{*}_{A}$ must belong to
$Q'$, or belong to $Q$ but not to $Q'$) leave us with the sole
option that, {\it once rephrased in the passive Hamiltonian
language}, the {\it active} diffeomorphism $D^{*}_{A}$ exploited
by the Argument must lie in the subset $Q'$ (${}_ADiff^{'}\ M^4$).
But, then, {\it it must necessarily map Cauchy data into
gauge-equivalent Cauchy data}, precisely those gauge-equivalent
data that generate the allegedly "different" solution within the
Hole. In the end, the "difference" turn out to correspond to a
mere different choice of the gauge for the same solution. Thus
{\it Leibniz equivalence} boils down to mere {\it gauge
equivalence} in its strict sense\footnote{The physical meaning of
this equivalence will be clarified in Section IV.}, an effect that
- for what said above - cannot be transparently displayed in the
configurational geometric description. On the other hand, were the
{\it active} diffeomorphism $D^{*}_{A}$, once passively rephrased,
to belong to the group $Q$ but not to the subset $Q^{'}$ (i.e.,
were it originally lying in ${}_ADiff\, M^4 $, but not in
${}_ADiff^{'}\, M^4)$), then it would not correspond to a mere
gauge equivalence and it would necessarily modify the Cauchy data
outside the Hole. Therefore it would lead to a really different
Einstein "universe" but it would violate the assumption of the
Hole Argument that $D^{*}_{A}$ be the identity on the Cauchy
hyper-surface. In any case, it is seen that the disappearance of
the "indeterminacy" rests upon the necessity of formulating the
Cauchy problem {\it before} talking about the relevant properties
of the solutions.

We conclude that - to the extent that the Cauchy problem is
well-posed, i.e. {\it in every globally hyperbolic space-time} and
not necessarily in our class only - exploiting the original Hole
Argument to the effect of asking ontological questions about the
general relativistic space-time is an enterprise devoid of real
philosophical impact, in particular concerning the menace of
indeterminism. There is clearly no room left for upholding
manifold substantivalism, "different worlds", "metric
essentialism" or any other metaphysical doctrine about space-time
points {\it in the face of the Hole Argument}. Of course, such
metaphysical doctrines can still be defended, yet independently of
the Hole story. The Hole Argument maintains nevertheless an
interesting open question regarding the issue of the physical
(viz. dynamical) individuation of the point-events of $M^4{}{}{}$
(see Section V).

\section{The Christodoulou-Klainermann
space-times, 3+1 splitting, and ADM canonical reduction}

\noindent The Christodoulou-Klainermann space-times are a
continuous family of space-times that are non-compact, globally
hyperbolic, asymptotically flat at spatial infinity (asymptotic
Minkowski metric, with asymptotic Poincar\'e symmetry group) and
topologically trivial ($M^4 \equiv R^3 \times R$), supporting
global 4-coordinate systems.

The ADM  Hamil\-to\-nian ap\-pro\-ach  starts with a 3+1 splitting
of the 4-dimensional manifold $M^4$ into constant-time
hyper-surfaces $\Sigma_{\tau} \equiv R^3$, indexed by the {\it
parameter time} $\tau$, each equipped with coordinates $\sigma^a$
(a = 1,2,3) and a three-metric ${}^3g$ (in components
${}^3g_{ab}$). The {\it parameter time} $\tau$ and the coordinates
$\sigma^a$ (a = 1,2,3) are in fact {\it Lorentz-scalar, radar}
coordinates {\it adapted} to the 3+1 splitting (Alba \& Lusanna,
2003, 2005a). They are defined with respect to an arbitrary
observer, a centroid $X^{\mu}(\tau )$, chosen as origin, whose
proper time may be used as the parameter $\tau$ labelling the
hyper-surfaces. On each hyper-surface all the clocks are
conventionally synchronized to the value $\tau$. Note that such
coordinates are intrinsically frame-dependent since they
parametrize a NIF centered on the arbitrary observer. The
simultaneity (and Cauchy) hyper-surfaces $\Sigma_{\tau}$ are
described by the embedding functions $x^{\mu} = z^{\mu}(\tau ,
\sigma^a) = X^{\mu}(\tau ) + F^{\mu}(\tau , \sigma^a)$,
$F^{\mu}(\tau , 0^a) = 0$ \footnote{Let us stress that the {\it
radar-coordinates} are not ordinary coordinates $x^{\mu}$ in a
chart of the Atlas ${\cal A}$ of $M^4$. They should be properly
called {\it pseudo-coordinates} in a chart of the Atlas ${\cal
\tilde A}$ defined by adding to $M^4$ the extra-structure of all
its admissible 3+1 splittings: actually the new coordinates are
{\it adapted} to this extra-structure. If the embedding of the
constant-time hyper-surfaces $\displaystyle \Sigma_{\tau}$ of a
3+1 splitting into $M^4$ is described by the functions
$\displaystyle z^{\mu}(\tau ,\sigma^a)$, then the transition
functions from the {\it adapted radar-coordinates} $\sigma^A =
(\tau; \sigma^a)$ to the ordinary coordinates are
$\frac{\displaystyle \partial z^\mu(\tau, \sigma^a)}{\displaystyle
\partial \sigma^A}$.}.

An important point to be kept in mind is that the explicit
functional form of embedding functions and - consequently - of the
geometry of the 3 + 1 splitting of $M^4$, thought to be implicitly
given at the outset, remains arbitrary until the solution of
Einstein's equations is worked out in a fixed gauge: see later.

\medskip

Now, start at a point on $\Sigma_{\tau}$, and displace it
infinitesimally in a direction that is normal to $\Sigma_{\tau}$.
The resulting change in $\tau$ can be written as $\triangle\,
\tau$ = $N d\tau$, where N is the so-called {\it lapse function}.
Moreover, the time displacement $d\tau$ will also shift the
spatial coordinates: $\sigma^{a}(\tau + d\tau)$ = $\sigma^a(\tau)+
N^{a}d\tau$, where $N^a$ is the {\it shift vector}. Then the
interval between $(\tau,\sigma^{a})$ and $(\tau + d\tau,
\sigma^{a} + d\sigma^{a})$ is: $ds^2 = N^2d\tau^2 -
{}^3g_{ab}(d\sigma^a + N^a d\tau)(d\sigma^b + N^b d\tau)$. The
{\it configurational} variables $N$, $N^{a}$, ${}^3g_{ab}$
(replacing the 4-metric $g$) together with their 10 conjugate
momenta, index a 20-dimensional phase space\footnote{Of course,
all these {\it variables} are in fact {\it fields}.}. Expressed
({\it modulo} surface terms) in terms of the ADM variables, the
Einstein-Hilbert action is a function of $N$, $N^{a}$,
${}^3g_{ab}$ and their first time-derivatives, or equivalently of
$N$, $N^{a}$, ${}^3g_{ab}$ and the extrinsic curvature
${}^3K_{ab}$ of the hyper-surface $\Sigma_{\tau}$, considered as
an embedded manifold.

Since Einstein's original equations are not hyperbolic, it turns
out that the canonical momenta are not all functionally
independent, but satisfy four conditions known as {\it primary}
constraints (they are given by the vanishing of the {\it lapse}
and {\it shift} canonical momenta). Another four, {\it secondary}
constraints, arise when we require that the primary constraints be
preserved through evolution (the secondary constraints are called
the {\it super-hamiltonian} $\mathcal{H}_0 \approx 0 $, and the
{\it super-momentum} $\mathcal{H}_a \approx 0, \ (a = 1,2,3)$
constraints, respectively). The eight constraints are given as
functions of the canonical variables that vanish on the {\it
constraint surface}. The existence of such constraints implies
that not all the points of the 20-dimensional phase space
represent physically meaningful states: rather, we are restricted
to the {\it constraint surface} where all the constraints are
satisfied, i.e., to a 12-dimensional (20 - 8) {\it surface} which,
however, does not possess the geometrical structure of a true
phase space. When used as generators of canonical transformations,
the eight constraints map points on the constraint surface to
points on the same surface; these transformations are known as
{\it gauge transformations}.

To obtain the correct dynamics for the constrained system, we must
consider the Dirac Hamiltonian, which is the sum of the DeWitt
surface term (DeWitt, 1967) \footnote{The DeWitt surface term is
{\it uniquely} determined as the sum of two parts: a) the surface
integral to be extracted from the Einstein-Hilbert action to get
the ADM action; b) a surface integral due to an integration by
parts required by the Legendre transformation from the ADM action
to phase space [see (Lusanna, 2001) after Eq.(5.5) and (Hawking \&
Horowitz, 1996)]. By adding a surface term different from the ADM
one, we would get another action with the same equations of motion
but an a-priori different canonical formulation. Still another
option is to consider the metric and the Christoffel connection as
independent configuration variables: this is the first-order
Palatini formalism, which has a much larger gauge freedom
including also second class constraints. All these canonical
formulations must lead anyway to the same number of physical
degrees of freedom.} [present only in spatially non-compact
space-times and becoming the ADM energy after suitable
manipulations (Lusanna, 2001; DePietri \& Lusanna \& Martucci \&
Russo, 2002)], of the {\it secondary} constraints multiplied by
the lapse and shift functions, and of the {\it primary}
constraints multiplied by arbitrary functions (the so-called {\it
Dirac multipliers}). If, following Dirac, we make the reasonable
demand that the evolution of all {\it physical variables} be
unique - otherwise we would have real physical variables that are
indeterminate and therefore neither {\it observable} nor {\it
measurable} - then the points of the constraint surface lying on
the same {\it gauge orbit}, i.e. linked by gauge transformations,
must describe the {\it same physical state}. Conversely, only the
functions in phase space that are invariant with respect to gauge
transformations can describe physical quantities.

To eliminate this ambiguity and create a one-to-one mapping
between points in the phase space and physical states, we must
impose further constraints, known as {\it gauge conditions} or
{\it gauge-fixings}. The gauge-fixings can be implemented by
arbitrary functions of the canonical variables, except that they
must define a {\it reduced phase space} that intersects each gauge
orbit exactly once ({\it orbit conditions}). The number of
independent gauge-fixing must be equal to the number of
independent constraints (i.e. 8 in our case). The canonical
reduction follows a cascade procedure\footnote{This procedure is
the natural one for systems with first-class constraints, because
it avoids mathematical inconsistencies like instabilities in the
pre-symplectic geometric structure. Usually one adds the gauge
fixings to all the constraints without taking this point into
consideration, so that, especially in GR, there is the possibility
to get coordinate singularities in a finite time.}. Precisely, the
gauge-fixings to the {\it super-hamiltonian} and {\it
super-momentum} come first (call it $\Gamma_4$): they determine
the 3-coordinate system and the {\it off-shell} shape of
$\Sigma_{\tau}$; then the requirement of their time constancy
fixes the gauges with respect to the {\it primary} constraints:
they determine the {\it lapse} and {\it shift} functions. Finally
the requirement of time constancy for these latter gauge-fixings
determines the Dirac multipliers. Therefore, the first level of
gauge-fixing gives rise to a {\it complete} gauge-fixing, say
$\Gamma_8$, and is sufficient to remove all the gauge
arbitrariness. This {\it is equivalent to the choice of a NIF},
namely, as said in footnote 7, to the determination of the {\it
appearances} of phenomena in a global extended laboratory (see
later) .

\medskip

The $\Gamma_8$ procedure reduces the original 20-dimensional phase
space to a copy ${\Omega}_4$ of the abstract {\it reduced
phase-space} $\tilde{\Omega}_4$ having 4 degrees of freedom per
point (12 - 8 gauge-fixings). Abstractly, the reduced phase-space
with its symplectic structure is defined by the quotient of the
constraint surface with respect to the 8-dimensional group of
gauge transformations and represents {\it the space of the
abstract gauge-invariant observables of GR: two configurational
and two momentum variables}.  These observables carry the physical
content of the theory in that they represent the {\it intrinsic
degrees of freedom of the gravitational field} (remember that at
this stage we are dealing with a pure gravitational field without
matter).

For any complete gauge $\Gamma_8$, we get a $\Gamma_8$-dependent
{\it copy} $\Omega_4$ of the abstract $\tilde{\Omega}_4$ in terms
of the symplectic structure (Dirac brackets) defined by the given
gauge-fixings and coordinatized by four {\it Dirac observables}
[call such field observables $q^r$, $p_s$ (r,s = 1,2)]. The
functional form of these Dirac observables ({\it concrete
realization of the gauge-invariant abstract observables in the
given complete gauge $\Gamma_8$}) in terms of the original
canonical variables depends upon the chosen gauge, so that such
observables - a priori - are neither tensors nor invariant under
${}_PDiff$. In each gauge $\Gamma_8$, the original 8 gauge
variables are now uniquely determined functions of the Dirac
observables. Yet, {\it off shell}, barring sophisticated
mathematical complications, {\it any two copies of $\Omega_4$ are
diffeomorphic images of one-another}.

\medskip

It is important to understand qualitatively the geometric meaning
of the eight infinitesimal {\it off-shell} Hamiltonian gauge
transformations and thereby the geometric significance of the
related gauge-fixings. i) The transformations generated by the
four {\it primary} constraints modify the {\it lapse} and {\it
shift} functions which, in turn, determine both how densely the
space-like hyper-surfaces $\Sigma_{\tau}$ are distributed in
space-time and the appearance of {\it gravito-magnetism} on them;
ii) the transformations generated by the three {\it
super-mo\-men\-tum} constraints induce a transition on
$\Sigma_{\tau}$ from one given 3-coordinate system to another;
iii) the transformation generated by the {\it super-hamiltonian}
constraint induces a transition from one given {\it a-priori}
"form" of the 3+1 splitting of $M^4$ to another (namely, from a
given notion of distant simultaneity to another), by operating
deformations of the space-like hyper-surfaces in the normal
direction.

It should be stressed that the {\it manifest effect} of the
gauge-fixings related to the above transformations emerges only
{\it at the end} of the canonical reduction and {\it after} the
solution of the Einstein-Hamilton equations has been worked out
(i.e., {\it on shell}). This happens because the role of the
gauge-fixings is essentially that of choosing the {\it functional
form} in which all the gauge variables depend upon the {\it Dirac
observables}, i.e. - physically - of fixing the form of the {\it
inertial potentials} of the associated NIF. As anticipated in the
Introduction, this important physical aspect is completely lost
within the abstract reduced phase space $\tilde{\Omega}_4$, which
could play, nevertheless, another role (see Sections V and VI).

It is only after the initial conditions for the {\it Dirac
observables} have been arbitrarily selected on a Cauchy surface
that the whole four-dimensional chrono-geometry of the resulting
{\it Einstein "universe"} is {\it dynamically} determined,
including the embedding functions $x^{\mu} = z^{\mu}(\tau ,\vec
\sigma)$ (i.e. the {\it on-shell} shape of $\Sigma_{\tau}$). In
particular, since the transformations generated by the
super-hamiltonian modify the rules for the synchronization of
distant clocks, {\it all} the relativistic {\it conventions},
associated to the 3 + 1 slicing of $M^4$ in a given {\it Einstein
"universe"}, turn out to be {\it dynamically-determined,
gauge-related options}\footnote{Unlike the special relativistic
case where the various possible conventions are non-dynamical
options.}.

\bigskip

\noindent Two important points must be emphasized.

First, in order to carry out the canonical reduction {\it
explicitly}, before implementing the gauge-fixings we have to
perform a basic canonical transformation at the {\it off-shell}
level, the so-called Shanmugadhasan transformation
(Shanmugadhasan, 1973; Lusanna, 1993), moving from the original
canonical variables to a new basis including the Dirac observables
as a canonical subset\footnote{In practice, this transformation is
adapted to seven of the eight constraints (Lusanna, 2001; DePietri
\& Lusanna \& Martucci \& Russo, 2002): they are replaced by seven
of the new momenta whose conjugate configuration variables are the
gauge variables describing the {\it lapse} and {\it shift}
functions and the choice of the spatial coordinates on the
simultaneity surfaces. The new basis contains the conformal factor
(or the determinant) of the 3-metric, which is determined by the
super-hamiltonian constraint (though as yet no solution has been
found for this equation, also called the Lichnerowicz equation),
and its conjugate momentum (the last gauge variable whose
variation describes the normal deformations of the simultaneity
surfaces).}. It should be stressed here that it is not known
whether the Shanmugadhasan canonical transformation, and therefore
the GR observables, can be defined {\it globally} in Christodoulou
- Klainermann space-times. In most of the spatially compact
space-times this cannot be done for topological reasons. A further
problem is that in field theory in general the status of the
canonical transformations is still heuristic. Therefore the only
tool (viz. the Shanmugadhasan transformation) we have for a
systematic search of GR observables in every type of space-time is
still lacking a rigorous definition. In conclusion, {\it the
mathematical basis of our analysis regarding the objectivity of
points is admittedly heuristic, yet our arguments are certainly no
more heuristic than the overwhelming majority of the theoretical
and/or philosophical claims concerning every model of GR}.

\medskip

\noindent The Shanmugadhasan transformation is highly {\it
non-local} in the metric and curvature variables: although, at the
end, for any $\tau$, the Dirac observables are {\it fields}
indexed by the coordinate point $\sigma^a$, they are in fact {\it
highly non-local functionals of the metric and the extrinsic
curvature over the whole {\it off shell} surface $\Sigma_{\tau}$}.
We can write, {\it symbolically}:

\begin{eqnarray}
q^r(\tau, \vec\sigma)
   &=& {{\mathcal{F}}_{[\Sigma_{\tau}]}}^r
  \bigr[(\tau, \vec\sigma)|\ {}^3g_{ab},
  {}^3{\pi}^{cd}\bigl]
 \nonumber \\
p_s(\tau, \vec\sigma)
  &=& {{\mathcal{G}}_{[\Sigma_{\tau}]}}_s
    \bigr[(\tau, \vec\sigma)|\ {}^3g_{ab},
    {}^3{\pi^{cd}}\bigl], \quad r,s = 1,2.
 \label{E4}
\end{eqnarray}

\medskip

Second: since, as mentioned, in {\it spatially compact}
space-times the original canonical Hamiltonian in terms of the ADM
variables is zero, the Dirac Hamiltonian happens to be written
solely in terms of the eight constraints and Lagrangian
multipliers. This means, however, that this Hamiltonian generates
purely harmless gauge transformations, so that {\it it cannot
engender any real temporal change}. Therefore, in
spatially-compact space-times, in a completely fixed Hamiltonian
gauge we have a vanishing Hamiltonian, and the canonical Dirac
observables are constant of the motion, i.e. $\tau$-independent.

In these models of GR with {\it spatially-compact} space-times
without boundary (nothing is known if there is a boundary) there
is the problem of re-introducing the {\it appearance of evolution}
in a frozen picture. Without entering this debated topic [see the
viewpoints of Earman (2002,2003), Maudlin (2002), Rovelli
(1991,2002) as well as the criticisms of Kuchar (1992,1993) and
Unruh (1991)], we only add a remark on the {\it problem of time}.
In all the globally hyperbolic space-times (the only ones
admitting a canonical formulation) there is a mathematical time
$\tau$, labeling the simultaneity (and Cauchy) surfaces, which has
to be connected to some empirical notion of time (astronomical
ephemerides time, laboratory clock,...). In a GR model with the
frozen picture there is no physical Hamiltonian governing the
evolution in $\tau$ \footnote{Unless, following Kuchar (1993), one
states that the super-Hamiltonian constraint is not a generator of
gauge transformations but an effective Hamiltonian instead.} and
an open problem is how to define a local evolution in terms of a
clock built with GR observables (with a time monotonically
increasing with $\tau$) and how to parametrize other GR
observables in terms of this clock (see the evolving constants of
motion and the partial and complete observables of Rovelli (1991,
2002), as well as a lot of different point of views).
\medskip

Our advantage point, however, is that, in the case of {\it
spatially non-compact} space-times such as those we are dealing
with in this work, the generator of $\tau$-temporal evolution is
the {\it weak ADM energy}\footnote{The ADM energy is a Noether
constant of motion representing the {\it total mass} of the
instantaneous 3-"universe", just one among the ten asymptotic ADM
Poincare' {\it charges} that, due to the absence of
super-translations, are the only asymptotic symmetries existing in
Christodoulou-Klainermann space-times. Consequently, the Cauchy
surfaces $\Sigma_{\tau}$ must tend to space-like hyper-planes,
normal to the ADM momentum, at spatial infinity. This means that:
(i) such $\Sigma_{\tau}$'s are the {\it rest frame} of the
instantaneous 3-"universe"; (ii) asymptotic inertial observers
exist and have to be identified with the {\it fixed stars}, and
(iii) an asymptotic Minkowski metric is naturally defined. This
{\it asymptotic background} allows us to avoid a split of the
metric into a background metric plus a perturbation in the weak
field approximation (note that our space-times provide a model of
either the {\it solar system} or our {\it galaxy} but, as yet, not
a well-defined model for cosmology). Finally, if gravity is
switched off, the Christodoulou-Klainermann space-times collapse
to Minkowski space-time and the ADM Poincare' charges become the
Poincare' special relativistic generators. {\it These space-times
provide, therefore, the natural model of GR for incorporating
particle physics which, in every formulation, is a chapter of the
theory of representations of the Poincare' group on Minkowski
space-time in inertial frames, with the elementary particles
identified by the mass and spin invariants}. If we change the
boundary conditions, allowing the existence of super-translations,
the asymptotic ADM Poincare' group is enlarged to the
infinite-dimensional asymptotic SPI group (Wald, 1984) and we
loose the possibility of defining the spin invariant. Note that in
spatially compact space-times {\it with boundary} it could be
possible to define a boundary Poincare' group (lacking in absence
of boundary), but we know of no result about this case. The
mathematical background of these results can be found in (Lusanna,
2001; Lusanna \& Russo, 2002; DePietri \& Lusanna \& Martucci \&
Russo, 2002; Agresti \& DePietri \& Lusanna \& Martucci, 2004) and
references therein.}. Indeed, this quantity {\it does generate
real $\tau$-temporal modifications of the canonical variables},
which subsequently can be rephrased in terms of some empirical
clock monotonically increasing in $\tau$. It's important to stress
that the density ${\cal E}_{ADM}(\tau, \vec{\sigma})$ of the weak
ADM energy $\int d^3 \sigma {\cal E}_{ADM} (\tau, \vec{\sigma})$
is a {\it gauge-dependent quantity} since it contains the
potentials of the inertial forces explicitly. This is nothing else
than another aspect of the gauge-dependence problem of the energy
density in GR.

\medskip

Thus, the final Einstein-Dirac-Hamilton equations for the Dirac
observables are

\begin{equation}
\dot{q}^r = \{q^r, H_{\mathrm{ADM}}\}^*, \quad \dot{p}_s = \{p_s,
H_{\mathrm{ADM}}\}^*, \quad r,s = 1,2,
 \label{E5}
\end{equation}

\noindent where $H_{\mathrm{ADM}}$ is intended as the restriction
of the {\it weak ADM energy} to $\Omega_4$ and where the
$\{\cdot,\cdot\}^*$ are the Dirac brackets.
\medskip

In conclusion, within the Hamiltonian formulation, we found a
class of solutions in which - unlike what has been correctly
argued by Earman (Earman, 2002; Belot \& Earman, 1999, 2001) for
spatially- compact space-times - there is a {\it real,
NIF-dependent, temporal change}. But this of course also means
that the {\it frozen-time} picture, being model dependent, is not
a {\it typical} feature of GR.

On the other hand it is not clear whether the formulation of a
cosmological model for GR is necessarily limited to spatially
compact space-times without boundary. As already said, our model
is suited for the solar system or the galaxy. It cannot be
excluded, however, that our asymptotic inertial observers (till
now identified with the fixed stars) might be identified with the
preferred frame of the cosmic background radiation with our
4-metric including some pre-asymptotic cosmological term.

\section{Finding the last remnant of physical objectivity: the
intrinsic gauge and the dynamical individuation of point-events}

We know that only two of the ten components of the metric are
physically essential: it seems plausible then to suppose that only
this subset can act as an individuating field, and that the
remaining components play a different role.

Consider the following {\it four scalars invariant functionals}
(the eigenvalues of the Weyl tensor), written here in Petrov's
compressed notation:

\begin{eqnarray}
w_1 &=& \mathrm{Tr} \, (g W g W), \nonumber \\
w_2 &=& \mathrm{Tr} \, (g W \epsilon W), \nonumber \\
w_3 &=& \mathrm{Tr} \, (g W g W g W), \nonumber \\
w_4 &=& \mathrm{Tr} \, (g W g W \epsilon W),
 \label{E6}
\end{eqnarray}

\noindent where $g$ is the 4-metric, $W$ is the Weyl tensor, and
$\epsilon$ is the Levi--Civita totally antisymmetric tensor.

Bergmann and Komar (Komar, 1958; Bergmann \& Komar, 1960;
Bergmann, 1961, 1962) proposed a set of invariant {\it intrinsic
pseudo-coordinates} as four suitable functions of the $w_T$
\footnote{Modulo the equations of motion, the eigenvalues $w_T$
are functionals of the 4-metric and its first derivatives.},

\begin{equation}
\hat{I}^{[A]} = \hat{I}^{[A]} \bigr[ w_T[g(x),\partial g(x)]
\bigl], \quad A = 0,1,2,3.
 \label{E7}
\end{equation}

\noindent Indeed, under the hypothesis of no space-time
symmetries, the $\hat{I}^{[A]}$ can be used to label the
point-events of space-time, at least locally.\footnote{Problems
might arise if we try to extend the labels to the entire
space-time: for instance, the coordinates might turn out to be
multi-valued.} Since they are {\it scalars}, the $\hat{I}^{[A]}$
are invariant under passive diffeomorphisms (therefore they do not
define a coordinate chart in the usual sense, precisely as it
happens with {\it radar} coordinates).

Clearly, our attempt to use intrinsic coordinates to provide a
physical individuation of point-events would {\it prima facie}
fail in the presence of symmetries, when the $\hat{I}^{[A]}$
become degenerate. This objection was originally raised by Norton
(see Norton, 1988, p.60) as a critique to
manifold-plus-further-structure (MPFS) substantivalism (see for
instance Maudlin, 1988, 1990). Several responses are possible.
First, although to this day all the {\it known} exact solutions of
the Einstein equations admit one or more symmetries, these
mathematical models are very idealized and simplified; in a
realistic situation (for instance, even with two masses alone)
space-time would be filled with the excitations of the
gravitational degrees of freedom, and would admit no symmetries at
all. Second, the parameters of the symmetry transformations can be
used as supplementary individuating fields, since, as noticed by
Stachel (1993), they also depend on metric field, through its
isometries. Third, and most important, in our analysis of the
physical individuation of points we are arguing a question of
principle, and therefore we must consider {\it generic} solutions
of the Einstein equations rather than the null-measure set of
solutions with symmetries.

\medskip
It turns out that the four Weyl scalar invariants can be
re-expressed in terms of the ADM variables, namely the {\it lapse}
$N$ and {\it shift} $N^a$ functions, the 3-metric ${}^3g_{ab}$ and
its conjugate canonical momentum (the extrinsic curvature
${}^3K_{a,b}$) \footnote{Bergmann and Komar have shown that the
four eigenvalues of the spatial part of the Weyl tensor depend
only upon the 3-metric and its conjugate momentum.}. Consequently
the $\hat{I}^{[A]}$ can be exploited to implement four
gauge-fixings constraints {\it involving a hyper-surface
$\Sigma_{\tau}$ and its embedding in $M^4$}. On the other hand, in
a completely fixed gauge $\Gamma_8$, the $\hat{I}^{[A]}$ become
{\it gauge dependent} functions of the Dirac observables of that
gauge.
\medskip

Writing

\begin{equation}
\hat{I}^{[A]} [w_T(g, \partial g)] \equiv \hat{Z}^{[A]} [{\hat
w}_T({}^3g, {}^3\pi, N, N^a)], \quad A = 0,1,2,3;
 \label{E8}
\end{equation}

\noindent and selecting a {\it completely arbitrary, radar,
pseudo-coordinate system} $\sigma^A \equiv [\tau,\sigma^a]$
adapted to the $\Sigma_\tau$ surfaces, we apply the {\it intrinsic
gauge-fixing} defined by

\begin{equation}
\chi^A \equiv \sigma^A - \hat{Z}^{[A]} \bigl[ {\hat
w}_T[{}^3g(\sigma^B), {}^3\pi(\sigma^D), N(\sigma^E),
N^a(\sigma^F)] \bigr] \approx 0, \quad A, B, D, E, F = 0,1,2,3;
 \label{E9}
\end{equation}

\noindent to the {\it super-hamiltonian} (A = 0) and the {\it
super-mo\-men\-tum} (A = 1,2,3) constraints. This is a good
gauge-fixing provided that the functions $\hat{Z}^{[A]}$ are
chosen to satisfy the fundamental {\it orbit conditions}
$\{\hat{Z}^{[A]},\mathcal{H}_B\} \neq 0, \quad(A,B = 0,1,2,3)$,
which ensure the independence of the $\chi^A$ and {\it carry
information about the Lorentz signature}. Then the complete
$\Gamma_8$ {\it intrinsic gauge-fixing} leads to

\begin{equation}
\sigma^A \equiv \tilde{Z}^{[A]} [ q^a(\sigma^B), p_b(\sigma^D)|
\Gamma)], \quad A, B, D = 0,1,2,3;\quad a,b = 1,2;
 \label{E10}
\end{equation}

\noindent where the notation indexed by $|\Gamma)$ means the
functional form assumed in the chosen gauge $\Gamma_8$.

The last equation becomes an {\it identity} with respect to the
$\sigma^A$, and amounts, {\it on-shell}, to a {\it definition} of
the {\it radar pseudo-coordinates} $\sigma^A$ as four {\it
scalars} providing a {\it physical individuation} of any
point--event, in the gauge-fixed coordinate system, in terms of
the gravitational degrees of freedom $q^a$ and $p_b$. In this way
each of the point--events of space-time is endowed with its own
{\it metrical fingerprint} extracted from the tensor field, i.e.,
the value of the four scalar functionals of the {\it Dirac
observables} (exactly four!)\footnote{The fact that there are just
{\it four} independent invariants for the vacuum gravitational
field should not be regarded as a coincidence. On the contrary, it
is crucial for the purpose of point individuation and for the
gauge-fixing procedure we are proposing.}. The price that we have
paid for this achievement is that we have broken general
covariance! This, however, is not a drawback because every choice
of 4-coordinates for a point (every gauge-fixing, in the
Hamiltonian language), in any procedure whatsoever for solving
Einstein's equations, amounts to a breaking of general covariance,
by definition. On the other hand the whole extent of general
covariance can be recovered by exploiting the gauge freedom.

Note that our construction does {\it not} depend on the selection
of a set of physically preferred intrinsic pseudo-coordinates,
because by modifying the functions $I^{[A]}$ we have the
possibility of implementing {\it any} (adapted) radar-coordinate
system. Passive diffeomorphism-invariance reappears in a different
suit: we find exactly the same functional freedom of ${}_PDiff\,
M^4$ in the functional freedom of the choice of the {\it
pseudo-coordinates} $Z^{[A]}$ (i.e., of the gauge-fixing). {\it
Any} adapted radar-coordinatization of the manifold can be seen as
embodying the physical individuation of points, because it can be
implemented as the Komar--Bergmann {\it intrinsic
pseudo-coordinates} after we choose the correct $Z^{[A]}$ and
select the proper gauge.

\medskip

In conclusion, as soon as the Einstein-Dirac-Hamilton equations
are solved {\it in the chosen gauge $\Gamma_8$}, starting from
given initial values of the {\it Dirac observables} on a Cauchy
hyper-surface $\Sigma_{\tau_{0}}$, the evolution in $\tau$
throughout $M^4$ of the Dirac observables themselves, whose
dependence on space (and on parameter time) is indexed by the
chosen coordinates $\sigma^A$, yields the following {\it
dynamically-determined} effects: i) {\it reproduces} the
$\sigma^A$ as the Bergmann-Komar {\it intrinsic
pseudo-coordinates}; ii) reconstructs space-time as an (on-shell)
foliation of $M^4$; iii) defines the associated NIF; iv)
determines a {\it simultaneity convention}.

Now, what happens if matter is present? Matter changes the Weyl
tensor through Einstein's equations and, in the new basis
constructed by the Shanmugadhasan transformation, contributes to
the separation of gauge variables from Dirac observables through
the presence of its own {\it Dirac observables}. In this case we
have {\it Dirac observables} for both the gravitational field and
the matter fields, which satisfy {\it coupled
Einstein-Dirac-Hamilton equations}. Since the gravitational Dirac
observables will still provide the individuating fields for
point-events according to our procedure, {\it matter will come to
influence the evolution of the gravitational Dirac observables and
thereby the physical individuation of point-events}. Of course, a
basic role of matter is the possibility of building apparatuses
for the measurement of the gravitational field, i.e. for an
empirical localization of point-events. As shown elsewhere (Pauri
\& Vallisneri, 2002, Lusanna \& Pauri, 2004a, 2004b), lacking a
dynamical theory of measurement, the epistemic circuit of GR can
be approximately closed via an {\it experimental three-steps
procedure} that, starting from concrete radar measurements and
using test-objects, ends up in a complete and empirically coherent
{\it intrinsic individuating gauge fixing}. In this way, the value
of the intrinsic coordinates at a point--event can be extracted
(in principle) by an actual experiment designed to measure the
$w_T$.

\medskip

Finally, let us emphasize that, even in the case with matter, time
evolution is still ruled by the {\it weak ADM energy}. Therefore,
the temporal variation corresponds to a {\it real change} and not
merely to a harmless gauge transformation as in other models of
GR. The latter include, as already stressed in Section IV, the
spatially compact space-time without boundary (or simply closed
models) which are exploited by Earman (2002). Since in these
spatially compact models the Dirac observables of every completely
fixed gauge are $\tau$-independent, the first of the gauge fixings
(\ref{E10}) is inconsistent: it is impossible to realize the {\it
time}-direction in terms of Dirac observables, and the
individuation of point-events breaks down. This is compatible with
the Wheeler-DeWitt interpretation according to which we can speak
only of a local time evolution (in the direction normal to
$\Sigma_{\tau}$) generated by the super-hamiltonian constraint
[see for instance Kuchar (1993)]: in other words the local
evolution would coincide with a continuous local change of the
convention about distant clock synchronization!

\section{Concluding remarks - I: the space-time physical texture}

The main results we have so far obtained are: i) a peculiar
dis-solution of the Hole Argument; ii) a NIF-dependent physical
individuation of point-events in terms of the intrinsic degrees of
freedom of the gravitational field (the essential {\it metrical
fingerprint} we were looking for); iii) a NIF-dependent temporal
evolution of the physical observables.

While the first claim is asserted for every globally-hyperbolic
space-time, the other results are valid with reference to the
particular spatially non-compact models of GR we have chosen. In
this Section we want to summarize the main features of these three
issues and devote a final Section to spend a few words about the
philosophical implications of our technical results, having in
view the traditional background of the substantivalism/relationism
debate as well as the recent debate on structural realism. It will
be seen that such implications instantiate a peculiar {\it
holistic and structuralist} view of the general-relativistic
space-time that we propose to call "point-structuralism".

Concerning the Hole Argument, our analysis of the correspondence
between symmetries of the Lagrangian configurational approach and
those of the Hamiltonian formulation has shown the following.
Solutions of Einstein's equations that, in the configurational
approach, differ within the Hole by elements of the subset
${}_ADiff^{'}\, M^4 $, which correspond to mappings among
gauge-equivalent Cauchy data, belong to the {\it same 4-geometry},
i.e. the same equivalence class generated by applying all {\it
passive diffeomorphisms} to any of the original 4-metrics: $
{}^4Geom  = {}^4Riem / {}_PDiff\, M^4 = {}^4Riem / Q'$.  In this
case, as seen at the Hamiltonian level, they are simply solutions
differing by a {\it harmless Hamiltonian gauge transformation on
shell} and describing, therefore, the same Einstein "universe".
Furthermore, it is possible to engender these allegedly different
solutions corresponding to the same "universe", by appropriate
choices of the initial gauge fixing (the functions
$\hat{Z}^{[A]}$). Since we know that the physical role of the
gauge-fixings is essentially that of choosing the functional form
of the inertial potentials in the NIF defined by the complete
gauge (the {\it epistemic} part of the game), the "differences"
among the solutions generated within the Hole by the allowed
active diffeomorphisms amount to the different {\it appearances}
of the intrinsic gravitational phenomena (the {\it ontic} part of
the game) in different NIFs. {\it In the end this is what,
physically, Leibniz equivalence reduces to}.

As already anticipated, our analysis contrasts with Stachel's
attitude about the Hole Argument. Leaving aside Stachel's broad
perspective about the significance and the possibility of
generalizations of the Hole story (see Stachel \& Iftime, 2005),
let us confine ourselves to few comments about the original
Stachel's proposal for the physical individuation of points of
$M^4$ by means of a fully covariant exploitation of the
Bergmann-Komar invariants $\hat{I}^{[A]} \bigr[ w_T[g(x),\partial
g(x)] \bigl], A = 0,1,2,3$. First of all, remember again that the
effect of the Hole Argument reveals itself on {\it solutions} of
Einstein equations and that the active diffeomorphisms that
purportedly maintain the physical identity of the points are,
therefore, dynamical symmetries. Now, how are we guaranteed that
the functional dependence of the quantities $\hat{I}^{[A]} \bigr[
w_T[g(x),\partial g(x)] \bigl]$ be {\it concretely characterized}
as relating to actual solutions of Einstein's equations ? Since in
the actual case we know that these quantities depend upon 4 Dirac
observables and 8 gauge viduate a solution, it follows that this
arbitrariness unavoidably transfers itself on the individuation
procedure and {\it leaves it undefined}. Indeed, speaking of
general covariance in an abstract way hides the necessity of
getting rid of the above arbitrariness by a gauge-fixing that, in
turn, necessarily breaks general covariance. In other words a
definite individuation entails a concrete characterization of the
{\it epistemic} part of the game, which is precisely what we have
done. The result is, in particular, exactly what Stachel's
suggestion was intended to, for our {\it intrinsic gauge} shows
that {\it active diffeomorphisms} of the first kind (i.e., those
belonging to $Q'$ in their passive interpretation) do map
individuations of point-events into {\it physically} equivalent
individuations. Indeed, since the {\it on-shell} Hamiltonian gauge
transformation connecting two different gauges is the passive
counterpart in $Q'$ of an {\it active} diffeomorphisms $D_A \in
{}_ADiff^{'}\, M^4$, it determines the {\it drag-along coordinate
transformation} ${\cal T}_{D_A}$ of Section II connecting the
radar 4-coordinates of the two gauges, i.e., the {\it dual view}
of the active diffeomorphism. While the active diffeomorphism {\it
carries-along} the identity of points by assumption, its passive
view attributes different physically-individuated
radar-coordinates to the same (mathematical) point. It is seen,
therefore, that for any point-event a given individuation by means
of Dirac observables is mapped into a physically-equivalent,
NIF-dependent individuation\footnote{Note that even the
description of two intersecting world-lines - {\it realizing a
point-coincidence} - is NIF-dependent !}.

As already noted, it's worth stressing again that the main reason
why we succeeded in carrying out a concrete realization of
Stachel's original suggestion to its natural end lies in the
possibility that the Hamiltonian method offers of working {\it
off-shell}. In fact, the 4-D active diffeomorphisms, {\it qua}
dynamical symmetries of Einstein's equations, must act on
solutions at every stage of the procedure and fail to display the
arbitrary {\it epistemic} part of the scalar invariants. On the
other hand, the Hamiltonian separation of the gauge variables
(characterizing the NIF and ruling the {\it generalized inertial
effects}), from the Dirac observables (characterizing tidal
effects) {\it is an off-shell procedure} that brings the wanted
{\it metrical fingerprint} by working independently of the initial
value problem. Once again, this mechanism is a typical consequence
of the special role played by gauge variables in GR\footnote{We
noted already that, according to a {\it main conjecture} advanced
elsewhere (Lusanna \& Pauri, 2004a, 2004b), a canonical basis
should exist having an explicit {\it scalar} character. An
evaluation of the degrees of freedom in connection with the
Newman-Penrose formalism for tetrad gravity (Stewart, 1993) tends
to corroborate the conjecture. In the Newman-Penrose formalism we
can define ten coordinate-independent quantities, namely the ten
Weyl scalars. If we add ten further scalars built using the
extrinsic curvature, we have a total of twenty scalars from which
one should extract a canonical basis replacing the 4-metric and
its conjugate momenta. Consequently, it should be possible to find
{\it scalar} Dirac observables [the {\it Bergmann observables},
see Lusanna \& Pauri, 2004b] and {\it scalar} gauge variables
(were the task of finding a canonical basis of scalars too
difficult, a minimal requirement would be to characterize a
well-defined family of scalars closing an algebra under Poisson
Brackets). Then, the individuating functions of (\ref{E8}) would
depend on scalars only and the distinction between Dirac and gauge
observables would become {\it fully invariant}. Yet, the
gauge-fixing procedure would always break general covariance and
one should not forget, furthermore, that the concept of {\it
radar-coordinates} contains a built-in {\it frame-dependence} (see
Section IV). Finally, the energy density ${\cal E}_{ADM}(\tau,
\vec{\sigma})$ would remain a NIF-dependent quantity anyway.}.

\medskip

Concerning the physical individuation of point-events, what we got
in Section V is tantamount to claiming that the {\it physical role
of the gravitational field without matter} is exactly that of
individuating {\it physically} the points of $M^4$  as
point-events, by means of the four independent phase-space degrees
of freedom. As pointed out above, the mathematical structure of
the canonical transformation that separates the Dirac observables
from the gauge variables is such that the Dirac observables are
{\it highly non-local functionals of the metric and the extrinsic
curvature over the whole (off-shell) hyper-surface $\Sigma_\tau$}.
The same is clearly true for the {\it intrinsic
pseudo-coordinates} [see Eq.(\ref{E8})]. Since the extrinsic
curvature has to do with the embedding of the hyper-surface in
$M^4$, the Dirac observables do {\it involve geometrical elements
external to the Cauchy hyper-surface} itself. Furthermore, since
the temporal gauge (fixed by the scalar $Z^{[0]}$), refers to a
continuous interval of hyper-surfaces, the gauge-fixing identity
itself is {\it intrinsically four-dimensional}.

At this point we could even say that the existence of physical
{\it point-events} in our models of general relativity appears to
be synonymous with the existence of the Dirac observables for the
gravitational field. We advance accordingly the {\it ontological}
claim that - physically - {\it Einstein's vacuum space-time in our
models is literally identifiable with the autonomous degrees of
freedom of such a structural field}, while the specific
(NIF-dependent) functional form of the {\it intrinsic
pseudo-coordinates} associates such coordinates to the points of
$M^4$. The intrinsic gravitational degrees of freedom are - as it
were - {\it fully absorbed in the individuation of point-events}.
On the other hand, when matter is present, the individuation
methodology maintains its validity and shows how matter comes to
influence the physical individuation of point-events.

\medskip

We would like to surmise that the disclosure of the superfluous
structure hidden behind the Leibniz equivalence by means of the
physical individuation of point-events, renders even more glaring
the ontological diversity of the gravitational field with respect
to all other fields, even beyond its prominent causal role. It
seems substantially difficult to reconcile the nature of the
gravitational field with the standard approach of theories based
on a {\it background space-time} (to wit, string theory and
perturbative quantum gravity in general). Any attempt at
linearizing such theories unavoidably leads to looking at gravity
from the perspective of a spin-2 theory in which the graviton
stands at the same ontological level as other quanta. In the
standard approach of background-dependent theories of gravity,
photons, gluons and gravitons all live on the stage on an equal
footing. From the point of view set forth in this paper, however,
{\it non-linear gravitons} are at the same time both the stage and
the actors within the causal play of photons, gluons, and other
{\it material characters} such as electrons and quarks.

\medskip

Finally, let us remark that the class of spatially non-compact
models treated in this paper, even if not yet able to describe
cosmology \footnote{Let us stress in this connection that in
spatially compact cosmologies the use of particle physics
(essentially defined in non-compact space-times) for the
description of the, e.g., nucleosynthesis implies an {\it huge
extrapolation}. Basically, it is well-known that at the level of
quantum field theory in background curved space-times, a useful
particle interpretation of states does not, in general, even
exist.}, has been priviliged by taking into primary consideration
the fundamental issue of how to incorporate particle physics in
GR, at the classical level to start with. Classical string
theories and super-gravity theories include particles, but their
quantization requires the introduction of a background space-time
to define the particle Fock space. On the other hand the only well
developed form of background-independent quantum gravity (loop
quantum gravity), obtained by quantizing either the connection or
the loop representation of GR, leads to a quantum formulation
inequivalent to Fock space, so that till now it is not known how
to incorporate particle physics. We hope that our viewpoint,
taking into account the non-inertial aspects of GR, can be
developed to the extent to be able to reopen the program of
canonical quantization of gravity in a background independent way
by {\it quantizing the Dirac observables only}\footnote{See Alba
\& Lusanna, 2005b, for a preliminary attempt to define
relativistic and non-relativistic quantum mechanics in
non-inertial frames in Galilei and Minkowski space-times,
respectively.}. Note finally that the individuating relation
(\ref{E10}) is a numerical identity that has a {\it built-in
non-commutative structure}, deriving from the Dirac--Poisson
structure hidden in its right-hand side. The individuation
procedure transfers, as it were, the non-commutative Poisson-Dirac
structure of the Dirac observables onto the individuated
point-events, even though the coordinates on the l.h.s. of the
identity are c-number quantities. One could guess that such a
feature might deserve some attention in view of quantization, for
instance by maintaining that the identity, interpreted as a
relation connecting mean values, could still play some role at the
quantum level.

\medskip

A further interesting suggestion comes from the following passage
of Bergmann and Komar:

\begin{quotation}
{\footnotesize \noindent [...] in general relativity the identity
of a world point is not preserved under the theory's widest
invariance group. This assertion forms the basis for the
conjecture that some physical theory of the future may teach us
how to dispense with world points as the ultimate constituents of
space-time altogether. (Bergmann \& Komar, 1972, 27)}
\end{quotation}
Indeed, would it be possible to build a fundamental theory that is
grounded in the reduced phase space $\tilde{\Omega}_4$ of the
abstract gauge-invariant Dirac observables? This would be an {\it
abstract and highly non-local theory} of classical gravitation
that, transparency aside, would be stripped of all the {\it
epistemic machinery} (the gauge freedom) which is indispensable
for an empirical access to the theory. For any given Einstein's
"universe" with its topology, the abstract {\it Dirac fields} in
$\tilde{\Omega}_4$ are locally functions of the points $x$ of an
abstract mathematical manifold $\tilde{M}_4$ that is the
equivalence class of all our {\it concrete realizations} of
space-time, each one equipped with its gauge-dependent
individuation of points, NIF and inertial forces. Such fields must
be called {\it intrinsic} to the extent that they are no longer
NIF-dependent, and synthesize, as it were, the essential
properties of all the appearances shown by the gauges. Admittedly,
the global existence of $\tilde{\Omega}_4$ over $\tilde{M}_4$ is
subjected to a huge set of mathematical hypotheses which we will
not take into account here. Locally, however, the Dirac fields
certainly exist and we could introduce a coordinate system defined
by their values as intrinsic individuating system for the given
"universe".

Let us stress that, once Einstein's equations have been solved,
the metric tensor and all of its derived quantities, in particular
the light-cone structure, can be re-expressed in terms of {\it
Dirac observables} in a gauge-fixed functional form. Yet, if we
look at the reduction procedure the other way around, we could
imagine starting with a given choice of initial values for the
{\it Dirac observables} (i.e., the germ of a {\it "universe"}),
and adding all the required {\it gauge} variables as suitable
independent variables, so as to obtain at the end a space-time
expression for the {\it local field} $g_{\mu \nu}(x)$. Since the
relation between all tensor expressions and Dirac observables
depends on the gauge, the gauge freedom would represent also the
{\it flexibility} of the final local description of the {\it deep
non-local structure} of the theory, a local description that
supports the empirical access to the theory. In other words the
{\it gauge} structure could be seen as playing a crucial role in
the {\it re-construction} of the concrete spatiotemporal continuum
representation from a non-local structure. We see, therefore, that
even in the context of classical gravitational theory, the
spatiotemporal {\it continuum} plays the role of an {\it epistemic
precondition} of our sensible experience of macroscopic objects,
playing a role which is not too dissimilar from that enacted by
Minkowski {\it micro-space-time} in the local relativistic quantum
field theory (see Pauri, 2000). We believe that, from the
philosophical point of view, one could recognize much more
substance here than what could appear {\it prima facie} a simple
instantiation of the relationship between canonical structure and
locality that pervades contemporary theoretical physics.

Finally, can this basic freedom in the choice of the {\it local
realizations} be equated with a ``taking away from space and time
the last remnant of physical objectivity,'' as Einstein suggested?
We believe that, discounting Einstein's ``spatial obsession'' with
{\it realism as locality (and separability)}, a significant kind
of spatio-temporal objectivity survives. It is true that the {\it
functional form} of the Dirac observables is NIF-dependent; yet,
even leaving aside the role of the abstract phase space
$\tilde{\Omega}_4$, there is no a-priori {\it physical}
individuation of the manifold points independently of the metric
field, so we cannot say that the individuation procedures
corresponding to different gauges individuate {\it different}
point-events (see footnote 32). A {\it really different} physical
individuation should only be attributed to different initial
conditions for the {\it Dirac observables}, (i.e., to a different
{\it "universe"}). We can, therefore, say that general covariance
represents the horizon of {\it a priori} possibilities for the
physical appearance of space-time, possibilities that must be
actualized within any given solution of the dynamical equations in
terms of NIFs.

\bigskip

\section{Concluding remarks - II: An instantiation of
structural realism as "point-structuralism}.

We conclude spending a few words about the implications of our
results for some issues surrounding the recent debate on
scientific structural realism, as well as for the traditional
debate on the absolutist/relationist dichotomy.

As well-known, the term {\it scientific realism} has been
interpreted in a number of different ways within the literature on
philosophy of science, in connection with the progressive
sophistication of our understanding of scientific knowledge. Such
ways concern, e.g., realism about {\it observable or unobservable
entities}, and realism about {\it theories}. A further
ramifications of meanings has been introduced more recently by the
so called {\it structural realism} (the only attainable reality
are relations between (unobservable) objects), which originated a
division between the so-called {\it epistemic} structural realists
(entity realism is unwarranted) and the {\it ontic} structural
realists (the relations exhaust what exists), (see Simon, 2003).

From the logical point of view, we can assume that the concept of
{\it structure} refers to a (stable or not) set of {\it relations}
among a set of some kind of {\it constituents} that are put in
relations (the {\it relata}). The specification expressed by the
notion of {\it structural realism} introduces {\it some kind} of
ontological distinction between the role of the relations and that
of the constituents. At least two main exemplary possibilities
present themselves as obvious: (i) there are relations, in which
the constituents are (ontologically) primary and the relation
secondary; (ii) there are relations, in which the relation is
(ontologically) primary while the constituents are secondary, and
this even without any prejudice about the ultimate ontological
consistency of the constituents. In the case of {\it physical}
entities, one could cautiously recover in this connection the
traditional distinction between {\it essential} and {\it
non-essential} properties ({\it accidents}) in order to
characterize the {\it degree of (ontological) primacy} of the
relations versus the {\it relata} and vice versa (and this
independently of any metaphysical flavor possibly connected to the
above distinctions). For example, one could say that in the
extreme case (i) only {\it accidental} properties of the
constituents can depend upon the relational structure, while in
the extreme case (ii) at least one {\it essential} property of the
constituents depends upon the relational structure (saying that
{\it all} the essential properties of the relata depend upon the
relation would be tantamount to claiming that there exist only
relations without constituents, as the {\it ontic structural
realist} has it).

A further complication is connected to the nature of the structure
we are considering. For while at the logical level (leaving aside
the deep philosophical issue concerning the relationships between
mathematical structures and substances) the concept of {\it
mathematical structure} (e.g. a system of differential equations,
or even the bare mathematical manifold of point which provides the
first layer of our {\it representations} of the real space-time )
can be taken to be sufficiently clear for our purposes, the
definition of {\it physical structure} raises existential
philosophical problems immediately. For example, we believe that
it is very difficult to define a {\it physical structure} without
bringing in its {\it constituents}, and thereby granting them {\it
some kind of existence} and defending some sort of ({\it entity
realism}). Analogously, we believe that it is very difficult to
defend {\it structural realism} without also endorsing a {\it
theory realism} of some sort. However, both theses are not
universally shared.

Having said this, let us come back to the results we have obtained
in the previous Sections. The analysis based on our {\it intrinsic
gauge} has disclosed a remarkable and rich local structure of the
general-relativistic space-time for the considered models of GR.
In correspondence to every {\it intrinsic gauge} (\ref{E10}) we
found a gauge-related {\it physical individuation of point-events}
in terms of the Dirac observables of that gauge, i.e., in terms of
the {\it ontic} part of the gravitational field, as represented in
the clothes furnished by the gauge (by the NIF). Moreover, since
the gauge-fixing identity (\ref{E10}) is four-dimensional we have
an instantiation of {\it metrical holism} which, though local in
the temporal dimension and characterized by a {\it dynamical
stratification in 3-hyper surfaces}, is {\it four-dimensional}.

At this point we can assert that we have a kind of ontology in
which the identity of point-events is {\it conferred} upon them by
a complex relational structure in which they are holistically
enmeshed. This relational structure includes all the elements of
the {\it complete gauge fixing $\Gamma_8$}, summarized by a NIF,
and supported by a definite solution of Einstein's equations
throughout $M_4$, corresponding to given initial values for the
Dirac observables in that gauge (a definite Einstein "universe").
The identity of point-events, at this level, should be properly
termed as {\it gauge-objective}.

It seems, therefore, that we have disclosed a {\it holistic
structure} which is clearly {\it ontologically prior} to its {\it
constituents}, as to their physical identity, even if we cannot
agree with Cao's assertion (see Cao, 2003, p.111) that the
constituents, as mere {\it place-holders}, derive their meaning or
{\it even their existence} from their function and place in the
structure. Indeed, at any level of consideration of GR, the
practical level above all, one cannot avoid quantifying over
points, and we have just attributed a physical
meaning\footnote{Even operationally, in principle (see Pauri \&
Vallisneri, 2002; Lusanna \& Pauri, 2004a, 2004b, already quoted
in Section V).} to our radar-coordinate indexing of such points
which makes point-events as ontologically equivalent to the
existence of the gravitational field as an extended entity. Quite
in general, we cannot see how a place-holder can have any
ontological function in an evolving network of relationships
without possessing at least {\it some kind} of properties. Even
more, let us recover our previous claim that - physically -
Einstein's vacuum space-time in our models is literally {\it
identifiable with the autonomous degrees of freedom} of the
gravitational field in vacuum, and moreover that the intrinsic
gravitational degrees of freedom are - as it were - {\it fully
absorbed in the individuation of point-events}. Both conclusions
do in fact confer a sort of causal power to the
gravitationally-dressed points. Then, we can ask whether all has
already been said concerning the identity of points or whether
instead some kind of {\it intrinsic individuality} survives
beneath the variety of descriptions displayed by all the
gauge-related NIFs, and common to all these {\it appearances}. For
indeed this kind of {\it intrinsic identity} is just furnished by
the {\it abstract Dirac fields} residing within the phase-space
$\tilde{\Omega}_4$, which is nothing else that a quotient with
respect to all of the concrete realizations and appearances of the
NIFs. Accepting this view, we are led to a peculiar space-time
{\it structure} in which the relation/relata correspondence does
not fit with any of the extreme cases listed above, for one could
assert that while the abstract {\it essential} properties belong
to the constituents as seen in $\tilde{\Omega}_4$ (so that
abstract point-events in $M_4$ would be like - as it were - to
{\it natural kinds}), the totality of the physically concrete {\it
accidents} are displayed by means of the holistic relational
structure. This is the reason why we are proposing to call this
peculiar kind of space-time structuralism as {\it
point-structuralism}.

Summarizing, this view holds that space-time point-events (the
{\it relata}) do exist as individuals and we continue to quantify
over them; however, their properties can be viewed both as {\it
extrinsic} and relational, being conferred on them in a holistic
way by the whole structure of the metric field and the extrinsic
curvature on a simultaneity hyper-surface, and, at the same time
as {\it intrinsic}, being coincident with the autonomous degrees
of freedom of the gravitational field represented by the abstract
NIF-independent Dirac fields in $\tilde{\Omega}_4$. In this way
both the metric field and the point-events maintain their {\it own
manner of existence}, so that the structural texture of space-time
in our models does not force us to abandon an {\it entity realist}
stance about both the metric field and its points. We must,
therefore, refute the thesis according to which metrical relations
can exist without their constituents (the point-events).

\medskip

Concerning the traditional debate on the dichotomy
substantivalism/relationism, we believe that our analysis - as a
case study limited to the class of space-times dealt with - may
offer a {\it tertium-quid} solution to the debate by overcoming
it. First of all, let us recall that, in remarkable diversity with
respect to the traditional historical presentation of Newton's
absolutism {\it vis \'a vis} Leibniz's relationism, Newton had a
much deeper understanding of the nature of space and time. In two
well-known passages of {\it De Gravitatione}, Newton expounds what
could be defined an original {\it proto-structuralist view} of
space-time (see also Torretti (1987), and DiSalle (1994)). He
writes (our {\it emphasis}):

\begin{quotation}
{\footnotesize \noindent Perhaps now it is maybe expected that I
should define extension as substance or accident or else nothing
at all. But by no means, for it has {\it its own manner of
existence} which fits neither substance nor accidents [\ldots] The
parts of space derive their character from their positions, so
that if any two could change their positions, they would change
their character at the same time and each would be converted
numerically into the other {\it qua} individuals.  The parts of
duration and space are only understood to be the same as they
really are because of their mutual order and positions ({\it
propter solum ordinem et positiones inter se}); nor do they have
any other {\it principle of individuation} besides this order and
position which consequently cannot be altered. (Hall \& Hall,
1962, p.99, p.103.)}
\end{quotation}:

\noindent
On the other hand, in his relationist arguments, Leibniz
could exploit the principle of sufficient reason because Newtonian
space was {\it uniform}, as the following passage lucidly explains
(our {\it emphasis}):

\begin{quotation}

{\footnotesize \noindent Space being {\it uniform}, there can be
neither any {\it external} nor {\it internal reason}, by which to
distinguish its parts, and to make any choice between them. For,
any external reason to discern between them, can only be grounded
upon some internal one. Otherwise we should discern what is
indiscernible, or choose without discerning. (Alexander (1956),
p.39).}
\end{quotation}

\noindent Clearly, if the parts of space were real, the Principle
of Sufficient Reason would be violated. Therefore, for Leibniz,
space is not real. The upshot, however, is that space (space-time)
in general relativity far from being {\it uniform} may possess, as
we have seen, a {\it rich structure}. This is just the reason why
- in our sense - it is {\it real}, and why Leibniz equivalence
called upon for general relativity happens to hide the very nature
of space-time, instead of disclosing it.

In conclusion, what emerges from our analysis is a kind of {\it
new structuralist} conception of space-time. Such new
structuralism is not only richer than that of Newton, as it could
be expected because of the dynamical structure of Einstein
space-time, but richer in an even deeper sense. For this {\it new
structuralist} conception turns out to include elements common to
the tradition of both {\it substantivalism} (space, and space
points, have an autonomous existence independently of other bodies
or matter fields) and {\it relationism} (the physical meaning of
space depends upon the relations between bodies or, in modern
language, the specific reality of space depends (also) upon the
(matter) fields it contains).

We have seen that the points of general-relativistic space-times,
quite unlike the points of the homogeneous Newtonian space, are
endowed with a remarkably rich {\it non-point-like} and {\it
holistic} structure furnished by the metric field and its
derivatives. Therefore, the independent degrees of freedom of the
metric field are able to characterize the "mutual order and
positions" of points {\it dynamically}, since - as it were - each
point-event "is" the "values" of the intrinsic degrees of freedom
of the gravitational field. This capacity is even stronger, since
{\it such mutual order is altered by the presence of matter}. On
the other hand, even though the metric field does {\it not} embody
the traditional notion of {\it substance}, it {\it exists} and
plays a role for the individuation of point-events by means of its
{\it structure}. On the other hand, although one can maintain also
the view that the physical properties are conferred to the
point-events in a peculiar relational form, our {\it
point-structuralism} does not support even the standard
relationist view. In fact, the holistic relationism we defend does
not reduce the whole of spatiotemporal relations to physical
relations (i.e. it is not eliminativist), nor it entails that
space-time does not exist as such, being reducible to physical
relations. It supports a thesis about the nature of identity of
point-events we continue to quantify over. These are individuals
in a peculiar sense: they exist as autonomous constituents, but
one cannot claim that their properties do not depend on the
properties of others. Not only relations do exist, but also the
carriers of them, even if they do bring intrinsic properties in a
very special sense.

We acknowledge that the validity of at least three of our results
is restricted to the class of models of GR we worked with. Yet, we
were interested in exemplifying a question of principle, so that
we can claim that there is a class of models of GR embodying both
a {\it real notion of NIF-dependent temporal change}, a {\it
NIF-dependent physical individuation of points} and a new {\it
structuralist and holistic view of space-time}.

\medskip

\section{Acknowledgments}

We thank our friend Mauro Dorato for many stimulating discussions

\vfill\eject

\section{References}

Alba,D. and Lusanna, L.(2003). Simultaneity, Radar 4-Coordinates
and the 3+1 Point of View about Accelerated Observers in Special
Relativity, submitted to General Relativity and Gravitation
(http://lanl.arxiv.org/abs/gr-qc/0311058). \medskip

Alba,D. and Lusanna, L. (2005a). Generalized Radar 4-Coordinates
and Equal-Time Cauchy Surfaces for Arbitrary Accelerated
Observers, (http://lanl.arxiv.org/abs/gr-qc/0501090). \medskip

Alba, D. and Lusanna, L. (2005b). Quantum mechanics in
non-inertial frames with a multi-temporal quantization scheme: I)
Relativistic particles (http://lanl.arxiv.org/abs/hep-th/0502060);
II) Non-relativistic Particles
(http://lanl.arxiv.org/abs/hep-th/0504060).

Agresti,J., De Pietri,R., Lusanna,L. and Martucci,L. (2004).
Hamiltonian Linearization of the Rest-Frame Instant Form of Tetrad
Gravity in a Completely Fixed 3-Orthogonal Gauge: a Radiation
Gauge for Background-Independent Gravitational Waves in a
Post-Minkowskian Einstein Space-Time, {\it General Relativity and
Gravitation} {\bf 36}, 1055-1134;
(http://lanl.arxiv.org/abs/gr-qc/0302084).\medskip .

Alexander,H (ed.),(1956).{\it The Leibniz-Clarke Correspondence},
Fourth Paper. Manchester: Manchester University Press. \medskip

Arnowitt,R., Deser,S., and Misner,C.W. (1962). The dynamics of
general relativity, in L. Witten (ed.), \emph{Gravitation: an
introduction to current research}, (pp. 227--265). NewYork: Wiley.
\medskip

Bartels,A. (1994). What is Spacetime if not a Substance?
Conclusions from the New Leibnizian Argument, in U. Mayer and
H.-J. Schmidt (eds.), \emph{Semantical Aspects of Spacetime
Theories}, (pp.41--51). Mannheim: B.I. Wissenshaftverlag.
\medskip

Belot,G. and Earman,J. (1999). From Metaphysics to Physics, in
J.Butterfield and C.Pagonis, (eds.), \emph{From Physics to
Philosophy}, (pp. 167--186). Cambridge: Cambridge University
Press.\medskip

Belot,G. and Earman,J.(2001). Pre-Socratic Quantum Gravity, in
C.Callender (ed.), {\it Physics Meets Philosophy at the Planck
Scale. Contemporary Theories in Quantum Gravity}. (pp. 213--255).
Cambridge: Cambridge University Press.\medskip

Bergmann, P.G. and Komar, A. (1960) Poisson Brackets between
Locally Defined Observables in General Relativity, {\it Physical
Review Letters} {\bf 4}, 432-433.\medskip

Bergmann, P.G. (1961) Observables in General Relativity, {\it
Review of Modern Physics} {\bf 33}, 510-514\medskip

Bergmann, P.G. (1962). The General Theory of Relativity, in
S.Flugge (ed.), {\it Handbuch derPhysik}, Vol. IV, {\it Principles
of Electrodynamics and Relativity}, (pp. 247-272). Berlin:
Springer.\medskip

Bergmann,P.G. and Komar, A. (1972), The Coordinate Group
Symmetries of General Relativity, {\it International Journal of
Theoretical Physics}, {\bf 5}, 15-28.\medskip

Butterfield,J. (1984). Relationism and Possible Worlds,
\emph{British Journal for the Philosophy of Science} \textbf{35},
1--13.\medskip

Butterfield,J.(1987) Substantivalism and Determinism,
\emph{International Studies in the Philosophy of Science}
\textbf{2}, 10--31.\medskip

Butterfield,J.(1988). Albert Einstein meets David Lewis, in A.
Fine and J. Leplin (eds.), \emph{PSA 1988}, \textbf{2}, (pp.
56--64).\medskip

Butterfield,J.(1989). The Hole Truth, \emph{British Journal for
the Philosophy of Science} \textbf{40}, 1--28.\medskip

Cao,T.(2003), Can we dissolve physical entities into mathematical
structure ?, {\it Synth\'ese}, \textbf{136}, 51-71\medskip

Christodoulou, D., and Klainerman, S. (1993). {\it The Global
Nonlinear Stability of the Minkowski Space}. Princeton: Princeton
University Press.\medskip

De Pietri,R., Lusanna,L., Martucci,L. and Russo,S. (2002). Dirac's
Observables for the Rest-Frame Instant Form of Tetrad Gravity in a
Completely Fixed 3-Orthogonal Gauge, {\it Geneneral Relativity and
Gravitation} {\bf 34}, 877-1033;
(http://lanl.arxiv.org/abs/gr-qc/0105084).\medskip

DeWitt,B. (1967) Quantum Theory of Gravity, I) The Canonical
Theory, {\it Physical Review} {\bf 160}, 1113-1148. II) The
Manifestly Covariant Theory, {\bf 162}, 1195-1239.\medskip

DiSalle,R. (1994) On Dynamics, Indiscernibility, and Spacetime
Ontology", {\it British Journal for the Philosophy of Science,
vol.45,}, pp.265-287.
\medskip

Dorato.M and Pauri,M. (2004) Holism and Structuralism in Classical
and Quantum General Relativity, Pittsburgh-Archive, ID code 1606,
forthcoming in (2006). S.French and D.Rickles (eds.), {\it
Structural Foundations of Quantum Gravity}, Oxford: Oxford
University Press.\medskip

Earman,J. and Norton,J. (1987). What Price Spacetime
Substantivalism? The Hole Story, \emph{British Journal for the
Philosophy of Science} \textbf{38}, 515--525.\medskip

Earman,J. (1989). \emph{World Enough and Space-Time}. Cambridge,
Mass.: The Mit Press.\medskip

Earman,J. (2002). Thoroughly Modern McTaggart or what McTaggart
would have said if He had read the General Theory of Relativity,
\emph{Philosophers' Imprint} \textbf{2}, No.3,
(http://www.philosophersimprint.org/002003/).\medskip

Earman, J. (2003). Ode to the constrained Hamiltonian formalism,
in K.Brading, and E.Castellani, (eds.), {\it Symmetries in
Physics: Philosophical Reflections}. Cambridge: Cambridge
University Press.\medskip

Einstein,A. (1914). Die formale Grundlage der allgemeinen
Relativit\"atstheorie, in \emph{Preuss. Akad. der Wiss. Sitz.},
(pp. 1030--1085).\medskip

Einstein,A. (1916). Die Grundlage der allgemeinen
Relativit\"atstheorie, \emph{Annalen der Physik} \textbf{49},
769--822; (1952) translation by W. Perrett and G. B. Jeffrey, The
Foundation of the General Theory of Relativity, in \emph{The
Principle of Relativity}, (pp. 117--118). New York: Dover.
\medskip

Friedman, M., (1983). {\it Foundations of Space-Time Theories}.
Princeton: Princeton University Press.
\medskip

Friedman, M.,(1984), {\it Roberto Torretti, Relativity and
Geometry}, Critical Review, \emph{No\^us} \textbf{18}, 653--664.
\medskip

Friedman, M.,(2001), Geometry as a Branch of Physics: Background
and Context for Einstein's "Geometry and Experience", in {\it
Reading Natural Philosophy: Essays in the History and Philosophy
of Science and Mathematics to Honor Howard Stein on His $70^{th}$
Birthday}, D.Malament (ed.). Chicago: Open Court.
\medskip

Friedrich, H. and Rendall, A. (2000). The Cauchy Problem for
Einstein Equations, in B.G.Schmidt (ed.), {\it Einstein's Field
Equations and their Physical Interpretation}. Berlin: Springer;
(http://lanl.arxiv.org/abs/gr-qc/0002074).\medskip

Hall, A.R. and Hall, M.B., (eds.), (1962). {\it De Gravitatione et
Aequipondio Fluidorum, Unpublished Scientific Papers of Isaac
Newton. A Selection from the Portsmouth Collection in the
University Library}. Canbridge: Cambridge University Press.
\medskip

Hawking, W.S. and Horowitz, G.T. (1996), The gravitational
Hamiltonian, action, entropy and surface terms, {\it Classical and
Quantum Gravity} {\bf 13}, 1487-1498.\medskip

Hilbert,D. (1917) Die Grundlagen der Physik. (Zweite Mitteilung),
\emph{Nachrichten von der K\"oniglichen Gesellschaft der
Wissenschaften zu G\"ottingen, Mathematisch-physikalische Klasse},
(pp. 53--76).\medskip

Komar,A. (1958). Construction of a Complete Set of Independent
Observables in the General Theory of Relativity, {\it Physical
Review} {\bf 111}, 1182-1187.\medskip

Kuchar, K. (1992). Time and interpretations of quantum gravity, in
{\it Proceedings of the 4th Canadian Conference on General
Relativity and Relativistic Astrophysics}, (pp. 211-314).
Singapore: World Scientific.\medskip

Kuchar, K. (1993). Canonical quantum gravity, 13th Conference on
General Relativity and Gravitation (GR-13), Cordoba, Argentina, 29
Jun - 4 Jul 1992. In G.Kunstatter, D.E.Vincent, and J.G.Williams
(eds.), {\it Cordoba 1992, General relativity and gravitation},
(pp.119-150);(http://lanl.arxiv.org/abs/gr-qc/9304012).\medskip

Lusanna, L. (1993). The Shanmugadhasan Canonical Transformation,
Function Groups and the Second Noether Theorem, {\it International
Journal of Modern Physics} {\bf A8}, 4193-4233.\medskip

Lusanna, L. (2001). The Rest-Frame Instant Form of Metric Gravity,
{\it General Relativity and Gravitation} {\bf 33}, 1579-1696;
(http://lanl.arxiv.org/abs/gr-qc/0101048).\medskip

Lusanna,L. and Russo,S. (2002) A New Parametrization for Tetrad
Gravity, {\it General Relativity and Gravitation} {\bf 34},
189-242; (http://lanl.arxiv.org/abs/gr-qc/0102074).\medskip

Lusanna, L. and M.Pauri, M. (2004a) The Physical Role of
Gravitational and Gauge Degrees of Freedom in General Relativity.
I: Dynamical Synchronization and Generalized Inertial Effects, to
appear in {\it General Relativity and Gravitation};
(http://lanl.arxiv.org/abs/gr-qc/0403081).\medskip

Lusanna,L. and Pauri,M. (2004b). The Physical Role of
Gravitational and Gauge Degrees of Freedom in General Relativity.
II: Dirac versus Bergmann Observables and the Objectivity of
Space-Time, to appear in {\it General Relativity and Gravitation};
(http://lanl.arxiv.org/abs/gr-qc/0407007).\medskip

Maudlin,T. (1988). The Essence of Space-Time, in \emph{PSA 1988},
\textbf{2}, (pp. 82--91).\medskip

Maudlin,T. (1990). Substances and Spacetimes: What Aristotle Would
Have Said to Einstein, \emph{Studies in the History and Philosophy
of Science} \textbf{21}, 531--61.\medskip

Maudlin, T. (2002). Throughly muddled McTaggart: or how to abuse
gauge freedom to generate metaphysical monstrosities. {\it
Philosophers' Imprint}, {\bf 2}, No. 4.\medskip

Norton,J. (1987). Einstein, the Hole Argument and the Reality of
Space, in J.Forge (ed.), {\it Measurement, Realism and
Objectivity}, Reidel, Dordrecht.\medskip

Norton,J.(1988). The Hole Argument, in PSA (1988), vol.2, pp.
56-64.\medskip

Norton,J.(1992). The Physical Content of General Covariance, in J.
Eisenstaedt, and A. Kox (eds.), {\it Studies in the History of
General Relativity}, Einstein Studies, vol. 3, (pp. 281--315).
Boston: Birkh\"auser.\medskip

Norton,J.(1993). General Covariance and the Foundations of General
Relativity: Eight Decades of Dispute, {\it Rep.Prog.Phys.} {\bf
56}, 791-858.\medskip

Pauri,M. (1996) Realt\`a e Oggettivit\`a, in F.Minazzi (ed.)
\emph{L'Oggettivit\`a nella Conoscenza Scientifica}, (pp.79--112).
Brescia: Franco Angeli.

\medskip

Pauri,M. (2000). Leibniz, Kant and the {\it Quantum}. In E.Agazzi
and M.Pauri (eds.), {\it The Reality of the Unobservable},
(pp.270-272). Boston Studies in the Philosophy of Science, N.215.
Dordrecht: Kluwer Academic Publishers.

\medskip

Pauri,M. and M.Vallisneri,M. (2002). Ephemeral Point-Events: is
there a Last Remnant of Physical Objectivity?, Essay in honor of
the 70th birthday of R.Torretti, {\it Dialogos} {\bf 79}, 263-303;
(http://lanl.arxiv.org/abs/gr-qc/0203014).\medskip

Rendall, A. (1998). Local and Global Existence Theorems for the
Einstein Equations, {\it Online Journal Living Reviews in
Relativity} {\bf 1}, n. 4; {\it ibid}. (2000) {\bf 3}, n. 1;
(http://lanl.arxiv.org/abs/gr-qc/0001008).\medskip

Rovelli, C. (1991). What is observable in classical and quantum
gravity. {\it Classical and Quantum Gravity}, {\bf 8},
297-316.\medskip

Rovelli, C. (2002). Partial observables. {\it Physical Review},
{\bf D, 65}, 124013, 1-8;
(http://lanl.arxiv.org/abs/gr-qc/0110035).\medskip

Rynasiewicz,R. (1994). The Lessons of the Hole Argument,
\emph{British Journal for the Philosophy of Science} \textbf{45},
407--436.\medskip

Rynasiewicz,R. (1996). Absolute versus Relational Space-Time: An
Outmoded Debate?, \emph{Journal of Philosophy} \textbf{43},
279--306.\medskip

Schlick,M. (1917) ({\it Raum und Zeit in der gegenw\"artigen
Physik}. Berlin: Springer; third edition 1920, fourth edition
1922; translated by H.Brose from third edition, {\it Space and
Time in Contemporary Physics} (Oxford: Oxford University Press,
1920; expanded to include changes in the fourth edition by
P.Heath, in H.Mulder and B. van de Velde-Schlick, (eds.), {\it
Moritz Schlick: Philosophical Papers}, vol.1. Dordrecht: Reidel,
(1978), pp.207-69.\medskip

Shanmugadhasan,S. (1973). Canonical Formalism for Degenerate
Lagrangians, {\it Journal of Mathematical Physics} {\bf 14},
677-687.\medskip

Simons,J. (ed).(2003). {\it Symposium on Structural Realism and
Quantum Field Theory}, {\it Synth\'ese} {\bf 136}. N.1.\medskip

Stachel,J. (1980). Einstein's Search for General Covariance,
1912--1915. Ninth International Conference on General Relativity
and Gravitation, Jena.\medskip

Stachel,J. (1986), in D.Howard and J.Stachel, (eds.),
\emph{Einstein and the History of General Relativity: Einstein
Studies}, vol. 1, (pp. 63--100). Boston: Birkh\"auser.\medskip

Stachel,J. (1993) The Meaning of General Covariance -- The Hole
Story, in J. Earman, I. Janis, G. J. Massey and N. Rescher,
(eds.), \emph{Philosophical Problems of the Internal and External
Worlds, Essays on the Philosophy of Adolf Gruenbaum},
(pp.129--160). Pittsburgh: University of Pittsburgh Press.
\medskip

Stachel,J. and Iftime,M. (2005) Fibered Manifolds, Natural
Bundles, Structured Sets, G-Sets and all that: The Hole Story from
Space Time to Elementary Particles, gr-qc/0505138.\medskip

Stewart,J. (1993). {\it Advanced General Relativity}, Cambridge:
Cambridge University Press.\medskip

Torretti,R. (1987). {\it Relativity and Geometry}, New York:
Dover, pp.167-168.\medskip

Wald,R.M. (1984) {\it General Relativity}. Chicago: University of
Chicago Press.\medskip

Weyl,H.,(1946). Groups, Klein's Erlangen Program. Quantities,
ch.I, sec.4 of {\it The Classical Groups, their Invariants and
Representations}, 2nd ed., (pp.13-23). Princeton: Princeton
University Press.

\end{document}